\documentclass[preprint,pteplogo]{ptephy_v2}
\preprintnumber{HUPD-2606} 

\usepackage{mathtools,bm}
\usepackage[hidelinks]{hyperref}
\usepackage[nameinlink,noabbrev]{cleveref}

\newcommand{\dd}{\mathrm{d}}
\newcommand{\sech}{\operatorname{sech}}
\newcommand{\vev}{v}
\newcommand{\Lzero}{L_0}
\newcommand{\Lam}{\Lambda}
\newcommand{\dlt}{\delta}
\newcommand{\Shat}{\widehat{S}_3}

\title{Localized Scalar Modes of $O(3)$ Critical Bubbles: Partial-Wave Continuum Mergers and Wave-Function Deformation}
\shorttitle{Localized Scalar Modes of $O(3)$ Critical Bubbles}
\shortauthorlist{T. Inagaki and Y. Murakami}

\author[1]{Tomohiro Inagaki$^{\ast}$}
\author[1]{Yuko Murakami}
\affil[1]{Information Media Center, Hiroshima University, Higashi-Hiroshima 739-8511, Japan
\email{inagaki@hiroshima-u.ac.jp}}

\begin{document}

\begin{abstract}

At phase coexistence, a degenerate quartic scalar potential admits an exact
planar kink whose normal fluctuation operator is the modified
P\"oschl--Teller operator, with translational and positive shape states below
a continuum beginning at $\Lam=4$.  We continue the two connected spectral
bands through finite supercooling in a smooth one-component quartic benchmark
and resolve $\ell=0,1,2$. The $O(3)$ bounce is obtained by singular
collocation, while the radial Euclidean Hessian is analyzed by
finite-difference diagonalization and independent threshold shooting. 
The
positive $\ell=2$ and $\ell=1$ branches reach the common false-vacuum continuum
threshold at $\dlt_{{\rm merge},2}=0.0900472$ and
$\dlt_{{\rm merge},1}=0.1410162$, respectively.  At each endpoint,
$u_\ell\propto\rho^{-\ell}$ is square integrable and defines a threshold
eigenstate; beyond the endpoint, however, no normalizable eigenstate continuation exists for the corresponding branch.
The $\ell=0$ shape state remains bound up to the geometric wall crossover. Its planar-mode overlap, radial centroid, and distinct interior and exterior decay
lengths reveal asymmetric wave-function deformation. 
Thus, angular spectral dissolution and the later geometric loss of a true-vacuum-like core are separate phenomena, revealing two distinct finite-supercooling fates of the planar shape mode.
\end{abstract}


\maketitle

\section{Introduction}
\label{sec:introduction}

First-order phase transitions proceed through the nucleation of bubbles of a
stable phase within a metastable background.  The semiclassical description is
organized around a Euclidean bounce and the spectrum of its quadratic
fluctuation operator \cite{Coleman1977,CallanColeman1977,Linde1983}.  At finite
temperature, the dominant saddle is commonly $O(3)$ symmetric, with one
negative mode associated with changes of the bubble radius and three
translational zero modes.  Beyond these universal directions, the wall may
support additional localized excitations whose properties depend on the shape
and curvature of the interface.  Resolving these modes state by state gives
information about wall deformations that is not contained in the aggregate
functional determinant.

Planar scalar walls provide a particularly transparent reference problem.  In
the quartic double-well theory, the exact kink reduces the normal fluctuation
equation to a modified P\"oschl--Teller problem
\cite{PoschlTeller1933,Dashen1974,Rajaraman1982}.  Besides the translational
mode, the spectrum contains a positive shape mode separated from the bulk
continuum.  Bound modes of extended walls have also been studied recently in
quantum corrections and Lorentzian worldvolume effective theories that couple
internal scalar modes to wall curvature and to the translational Goldstone
field \cite{Evslin2024,BlancoPillado2025}.

Positive discrete modes of finite-temperature bubbles have direct precedent.
Brihaye and Kunz constructed the discrete spectrum of spherically symmetric
electroweak bubbles, followed a positive mode with temperature, and showed that
it approaches the positive antikink mode near phase coexistence
\cite{BrihayeKunz1993}.  Related density-functional studies distinguished
shape and interfacial-width fluctuations in extended interfaces and calculated
capillary and higher-energy width modes for spherical and cylindrical
interfaces \cite{RobledoVarea1997,VareaRobledo1998}.  Iwamatsu and Okabe
solved the partial-wave fluctuation problem of critical bubbles in a
triple-parabolic square-gradient model and explicitly followed bound states
toward the continuum \cite{IwamatsuOkabe2010}.  These results motivate a
partial-wave-resolved continuation in a smooth field-theory potential.

For the three-dimensional critical bubble, the leading thin-wall deformation
of the P\"oschl--Teller spectrum is known analytically.  M\"unster and Rotsch
obtained both the displacement band and the positive band near eigenvalue
three, including the absence of a first-order eigenvalue correction
\cite{MuensterRotsch2000}; the low displacement branch has also been derived
in general dimension in modern thin-wall determinant calculations
\cite{Ivanov2022}.  Complementary work evaluates the complete one-loop
functional determinant rather than resolving individual positive states
\cite{BaackeKiselev1993,Kripfganz1995,DunneMin2005,BubbleDet2023,Matteini2025}.
The present work uses the known thin-wall spectrum as an analytic anchor and
then follows the corresponding localized states beyond the asymptotic
thin-wall regime.

For cosmological transitions, realistic bounce profiles are usually obtained
numerically from finite-temperature effective potentials.  Public tools now
compute bounce profiles and actions in single- and multi-field systems
\cite{Wainwright2012,Athron2019}, while current work emphasizes uncertainties
from supercooling, scale hierarchies, derivative expansions, and the coupling
of expanding walls to the thermal plasma
\cite{Athron2024,Kierkla2025,Ekstedt2025,Ai2025}.  A controlled single-field
benchmark is therefore useful for separating the spectral and spatial
evolution of localized modes from these additional model-dependent effects.

We study a smooth tilted quartic whose coexistence limit is the exactly
solvable planar P\"oschl--Teller problem.  The $O(3)$ bounce is obtained by
singular collocation, and the radial Euclidean Hessian is analyzed by
finite-difference diagonalization together with independent threshold
shooting.  Because the false-vacuum continuum threshold is common to all
partial waves, the positive angular states can be followed through their
square-integrable threshold endpoints.  For the surviving $\ell=0$ internal
state, we additionally track the normalized wave function, its overlap with
the planar shape mode, its radial centroid, and its interior and exterior
decay scales.

The central results are twofold.  First, threshold shooting determines the
partial-wave-dependent endpoints of the positive angular branches:
$\dlt_{{\rm merge},2}=0.0900472$ and
$\dlt_{{\rm merge},1}=0.1410162$.  At each endpoint the threshold state
remains square integrable, whereas no normalizable eigenstate continuation
exists beyond it.  Second, the $\ell=0$ shape state remains bound up to the
geometric wall crossover while its normalized profile becomes strongly
asymmetric.  The known near-coexistence coefficients are recovered as an
analytic and numerical consistency check, providing a controlled anchor for
the continuation to finite supercooling.  The angular branches and the
surviving radial branch thus realize two distinct finite-supercooling fates of
the planar shape mode.

The paper is organized as follows.  \Cref{sec:model} defines the tilted quartic
model and reviews the exact planar limit.  \Cref{sec:bubble} formulates the
spherical bounce and its thin-wall behavior.  \Cref{sec:fluctuations} gives the
radial fluctuation problem, and \cref{sec:numerics} describes the numerical
methods.  \Cref{sec:results} presents the bounce and spectral results, followed
by their interpretation and limitations in \cref{sec:discussion}.  We conclude
in \cref{sec:conclusion}.  Exact maps to standard quartic parametrizations and
numerical convergence checks are collected in the appendices.

\section{Tilted quartic model and exact planar limit}
\label{sec:model}

\subsection{From a finite-temperature effective potential to the reduced model}

For a slowly varying scalar order parameter, we approximate the equilibrium
finite-temperature effective action by the leading terms in its derivative
expansion,
\begin{equation}
 S_3[\phi;T]
 =
 \int \dd^3x\,
 \left[
 \frac{1}{2}Z(\phi,T)(\bm{\nabla}\phi)^2
 +V_{\rm eff}(\phi,T)
 +O(\bm{\nabla}^4)
 \right].
 \label{eq:thermal-effective-action}
\end{equation}
Here $Z(\phi,T)$ is the field-dependent wave-function renormalization.  In the
reduced benchmark studied below, we use a canonically normalized field and set
$Z=1$.  The finite-temperature potential may be obtained perturbatively with
the appropriate thermal resummations or by nonperturbative methods
\cite{DolanJackiw1974,ArnoldEspinosa1993,InagakiOgureSato1998,Athron2024}.
We use this framework only to motivate a reduced single-field quartic that
isolates vacuum nondegeneracy and spherical curvature while retaining the
exactly solvable coexistence limit.

Let $\phi_{\rm f}(T)$ and $\phi_{\rm t}(T)$ denote the false- and true-vacuum
minima of $V_{\rm eff}$, and define
\begin{equation}
 \Delta V(T)
 =
 V_{\rm eff}(\phi_{\rm f}(T),T)
 -
 V_{\rm eff}(\phi_{\rm t}(T),T).
 \label{eq:general-free-energy-difference}
\end{equation}
At the coexistence temperature $T_c$,
\begin{equation}
 \Delta V(T_c)=0,
 \label{eq:critical-degeneracy}
\end{equation}
so the two phases are degenerate.  If the potential between the two minima is
approximated by a quartic polynomial, degeneracy and stationarity imply the
factorized form
\begin{equation}
 V_{\rm eff}(\phi,T_c)-V_c
 \simeq
 \frac{\lambda_c}{4}
 \left(\phi-\phi_{{\rm f},c}\right)^2
 \left(\phi-\phi_{{\rm t},c}\right)^2,
 \label{eq:factorized-critical-potential}
\end{equation}
where $V_c$ is the common vacuum value and
$\phi_{{\rm f},c}\equiv\phi_{\rm f}(T_c)$,
$\phi_{{\rm t},c}\equiv\phi_{\rm t}(T_c)$.
For a quartic truncation this factorization is exact: the two degenerate
stationary minima are double zeros of $V_{\rm eff}-V_c$.

A familiar example is the high-temperature polynomial
\begin{equation}
 V_{\rm HT}(\phi,T)
 =
 D\left(T^2-T_0^2\right)\phi^2
 -ET\phi^3
 +\frac{\lambda_T}{4}\phi^4+\cdots,
 \label{eq:high-temperature-potential}
\end{equation}
which is widely used as an analytic representation of a thermally induced
first-order phase transition \cite{AndersonHall1992,ArnoldEspinosa1993}.  Here $D$
and $E$ are effective thermal coefficients, $T_0$ sets the quadratic thermal
scale, and $\lambda_T$ is the temperature-dependent quartic coupling; we use
$\lambda_c\equiv\lambda_{T_c}$ below.
If the false vacuum is at $\phi=0$ and the broken minimum at $T_c$ is
$\phi_c$, the critical stationarity and degeneracy conditions give
\begin{equation}
 \phi_c=\frac{2ET_c}{\lambda_c},
 \qquad
 D\left(T_c^2-T_0^2\right)
 =\frac{E^2T_c^2}{\lambda_c},
 \label{eq:high-temperature-critical-relations}
\end{equation}
and therefore
\begin{equation}
 V_{\rm HT}(\phi,T_c)-V_{\rm HT}(0,T_c)
 =
 \frac{\lambda_c}{4}\phi^2(\phi-\phi_c)^2.
 \label{eq:high-temperature-factorization}
\end{equation}
Thus the factorized quartic is not an arbitrary soliton potential; it is the
critical-temperature limit of a standard finite-temperature potential
exhibiting a first-order phase transition.

To describe a small departure from degeneracy, introduce the critical field
separation
\begin{equation}
 \vev
 =
 \phi_{{\rm t},c}-\phi_{{\rm f},c},
 \qquad
 s
 =
 \frac{\phi-\phi_{{\rm f},c}}{\vev}.
 \label{eq:field-rescaling}
\end{equation}
The lowest-degree polynomial that changes the relative vacuum energy while
leaving $s=0$ and $s=1$ stationary is the cubic Hermite interpolant
\begin{equation}
 P(s)=3s^2-2s^3,
 \qquad
 P(0)=0,\quad P(1)=1,\quad P'(0)=P'(1)=0.
 \label{eq:hermite-tilt}
\end{equation}
Accordingly, the near-coexistence potential may be decomposed as
\begin{equation}
 V_{\rm eff}(\phi,T)-V_{\rm eff}(\phi_{\rm f},T)
 =
 \frac{\lambda_c\vev^4}{4}
 \left[
 s^2(1-s)^2-\dlt(T)P(s)
 \right]
 +\Delta V_{\rm rem}(\phi,T),
 \label{eq:near-critical-decomposition}
\end{equation}
where
\begin{equation}
 \dlt(T)
 =
 \frac{4\Delta V(T)}{\lambda_c\vev^4}.
 \label{eq:delta-from-free-energy}
\end{equation}
Because $s$ is defined using the critical-temperature vacuum locations,
\cref{eq:near-critical-decomposition} is a projection in a fixed field
coordinate rather than an exact identity for a general microscopic
potential.  The remainder $\Delta V_{\rm rem}$ contains shifts of the vacuum
locations, temperature dependence of the quartic coupling, and higher-order
field dependence.  Our reduced model sets this remainder to zero, thereby
isolating vacuum nondegeneracy and spherical curvature while preserving the
exact P\"oschl--Teller reference problem at $T=T_c$.  
Quantitative matching to a microscopic model requires restoring these effects together with the associated relaxation of the bubble profile.

The thermodynamic meaning of $\dlt$ is particularly transparent close to
$T_c$.  If
\begin{equation}
 \mathcal{L}
 =
 T_c\left[
 \frac{\partial V_{\rm eff}(\phi_{\rm t},T)}{\partial T}
 -
 \frac{\partial V_{\rm eff}(\phi_{\rm f},T)}{\partial T}
 \right]
\end{equation}
is the latent heat density, then
\begin{equation}
 \Delta V(T)
 =
 \mathcal{L}\frac{T_c-T}{T_c}
 +O\!\left((T_c-T)^2\right),
 \label{eq:free-energy-near-critical}
\end{equation}
and hence
\begin{equation}
 \dlt(T)
 =
 \frac{4\mathcal{L}}{\lambda_c\vev^4}
 \frac{T_c-T}{T_c}
 +O\!\left(\frac{(T_c-T)^2}{T_c^2}\right).
 \label{eq:delta-supercooling-map}
\end{equation}
The parameter $\dlt$ is therefore a dimensionless measure of supercooling,
up to a model-dependent normalization fixed by the latent heat, critical field
separation, and quartic coupling.

\subsection{Reduced tilted quartic and the exact planar wall}
\label{sec:reduced-planar-wall}

After shifting the false vacuum to $\phi=0$ and writing
$\lambda\equiv\lambda_c$, the reduced potential is
\begin{equation}
 V(\phi;\dlt)
 =
 \frac{\lambda \vev^4}{4}
 \left[
 s^2(1-s)^2
 -\dlt\left(3s^2-2s^3\right)
 \right],
 \qquad
 s\equiv\frac{\phi}{\vev}.
 \label{eq:physical-potential}
\end{equation}
The interpolation polynomial has vanishing first derivative at $s=0$ and
$s=1$.  Consequently, these field values remain stationary for all $\dlt$.
Their potential difference is
\begin{equation}
 V(0;\dlt)-V(\vev;\dlt)
 =
 \frac{\lambda\vev^4}{4}\dlt.
 \label{eq:vacuum-difference}
\end{equation}
Thus $s=0$ is the false vacuum and $s=1$ the true vacuum for $\dlt>0$.
The false vacuum remains locally stable for
\begin{equation}
 0<\dlt<\frac{1}{3},
 \label{eq:delta-range}
\end{equation}
and the barrier lies at
\begin{equation}
 s_{\rm top}=\frac{1-3\dlt}{2}.
 \label{eq:barrier-position}
\end{equation}

We define the reference wall length
\begin{equation}
 \Lzero
 =
 \frac{2\sqrt{2}}{\vev\sqrt{\lambda}}.
 \label{eq:L0-definition}
\end{equation}
At degeneracy this scale coincides exactly with the slope-defined wall
thickness, as shown next.

At $\dlt=0$, the potential has two degenerate minima.  We denote by $z$ the
signed Cartesian distance normal to an infinite planar interface, with the
wall center placed at $z=0$; the two coordinates tangent to the wall are
suppressed because the background is translationally invariant along them.
For a large spherical bubble, this local normal coordinate is recovered as
$z\simeq r-R$, where $R$ is the bubble radius.  In terms of
$x=z/\Lzero$, the planar wall interpolating from the true phase at
$x\to-\infty$ to the false phase at $x\to+\infty$ is
\begin{equation}
 s_0(x)
 =
 \frac{1}{2}\left(1-\tanh x\right),
 \qquad
 \phi_0(z)=\vev\,s_0(z/\Lzero).
 \label{eq:planar-kink}
\end{equation}
Since $\max_x|s_0'(x)|=1/2$, the maximum-slope definition gives
\begin{equation}
 \frac{L_{\rm slope}^{(0)}}{\Lzero}
 =
 \frac{1}{2\max_x|s_0'(x)|}
 =1.
 \label{eq:planar-slope-width}
\end{equation}
Thus \(\Lzero\) is precisely the slope-defined thickness of the exact planar
wall.  This normalization will be used below to quantify the broadening of the
spherical bubble wall away from degeneracy.

For the degenerate kink, the first integral of the static field equation gives
$\tfrac12(\dd\phi_0/\dd z)^2=V(\phi_0;0)$.  The corresponding surface tension is
therefore
\begin{equation}
 \sigma_0
 =
 \int_{-\infty}^{\infty}\dd z
 \left(\frac{\dd\phi_0}{\dd z}\right)^2
 =
 \frac{\vev^2}{3\Lzero}.
 \label{eq:surface-tension}
\end{equation}

The Hessian is obtained by taking the second variation of the planar
free-energy functional about the kink background.  Defining
$V_{\phi\phi}\equiv\partial^2V/\partial\phi^2$ and using
$x=z/\Lzero$, its dimensionless wall-normal restriction is
\begin{align}
 \mathcal{M}_{\rm PT}
 &\equiv
 \Lzero^2
 \left[
 -\frac{\dd^2}{\dd z^2}
 +V_{\phi\phi}(\phi_0;0)
 \right]
 \nonumber\\
 &=
 -\frac{\dd^2}{\dd x^2}
 +4-6\sech^2x.
 \label{eq:PT-operator}
\end{align}
Here the factor of two follows from
$\Lzero^2\lambda\vev^2/4=2$, and the final line follows upon inserting
the exact kink profile in \cref{eq:planar-kink}.  Thus
$\mathcal M_{\rm PT}$ is the modified P\"oschl--Teller Hessian governing
fluctuations normal to the degenerate planar wall.
Its discrete eigenfunctions are orthonormalized according to
\begin{equation}
 \int_{-\infty}^{\infty}\dd x\,
 \chi_a^*(x)\chi_b(x)
 =
 \delta_{ab},
 \qquad
 a,b\in\{{\rm tr},{\rm sh}\},
 \label{eq:planar-mode-normalization}
\end{equation}
and may be chosen as
\begin{align}
 \chi_{\rm tr}(x)
 &=
 \sqrt{\frac{3}{4}}\sech^2x,
 &
 \Lam_{\rm tr}
 &=
 0,
 \label{eq:planar-translation}
 \\
 \chi_{\rm sh}(x)
 &=
 \sqrt{\frac{3}{2}}\sech x\tanh x,
 &
 \Lam_{\rm sh}
 &=
 3.
 \label{eq:planar-shape}
\end{align}
Here $\Lam_{\rm tr}$ and $\Lam_{\rm sh}$ are the corresponding discrete Hessian eigenvalues.  The continuous spectrum consists of generalized eigenfunctions labeled by the dimensionless wall-normal momentum $k\in\mathbb{R}$, with continuum eigenvalues
\begin{equation}
 \Lam_k=4+k^2.
 \label{eq:planar-continuum}
\end{equation}
Thus the continuum threshold is $\Lam=4$.
The exact spectrum in \cref{eq:planar-translation,eq:planar-shape,eq:planar-continuum} is the reference point for the numerical continuation performed below.

\section{Spherical critical bubble}
\label{sec:bubble}

At finite temperature, the dominant semiclassical saddle is commonly an
$O(3)$-symmetric critical bubble \cite{Linde1983}.  In the thin-wall limit its
surface and volume contributions balance,
$F(R)\simeq4\pi R^2\sigma-(4\pi/3)R^3\Delta V$, so the critical configuration
is a stationary Euclidean saddle with a large interface radius.

Here $r$ denotes the physical radial distance from the bubble center in
three-dimensional Euclidean space.
We introduce
\begin{equation}
 \rho=\frac{r}{\Lzero},
 \label{eq:dimensionless-radius}
\end{equation}
so that $\rho$ is the corresponding dimensionless radius.  The physical
bounce profile is
\begin{equation}
 \phi_b(r)=\vev\,s_b(r/\Lzero).
 \label{eq:physical-bounce-profile}
\end{equation}
In the large-radius limit, the local wall-normal coordinate introduced in
\cref{sec:reduced-planar-wall} is $z=r-R$.  The dimensionless bounce equation is
\begin{equation}
 s_b''(\rho)
 +\frac{2}{\rho}s_b'(\rho)
 =
 4s_b(1-s_b)(1-2s_b-3\dlt),
 \label{eq:bounce-equation}
\end{equation}
subject to
\begin{equation}
 s_b'(0)=0,
 \qquad
 s_b(\rho\to\infty)=0.
 \label{eq:bounce-boundaries}
\end{equation}
The dimensionless three-dimensional action is defined by
\begin{equation}
 \Shat(\dlt)
 \equiv
 \frac{S_3}{4\pi\vev^2\Lzero}
 =
 \int_0^\infty\dd\rho\,\rho^2
 \left[
 \frac{1}{2}(s_b')^2
 +2\widehat V(s_b;\dlt)
 \right],
 \label{eq:dimensionless-action}
\end{equation}
where
\begin{equation}
 \widehat V(s;\dlt)
 =s^2(1-s)^2-\dlt(3s^2-2s^3).
\end{equation}

We define the bubble radius by the half-height condition
\begin{equation}
 s_b(R_{1/2}/\Lzero)=\frac{1}{2},
 \label{eq:radius-definition}
\end{equation}
and the local wall width by the maximum slope,
\begin{equation}
 \frac{L_{\rm slope}}{\Lzero}
 =
 \frac{1}{2\max_\rho|s_b'(\rho)|}.
 \label{eq:width-definition}
\end{equation}
The normalization is chosen so that $L_{\rm slope}=\Lzero$ for the exact
planar wall at $\dlt=0$, as shown in \cref{eq:planar-slope-width}.

For $\dlt\ll1$, the bubble is thin walled.  
The leading expressions are
\begin{equation}
 \frac{R_{1/2}}{\Lzero}
 \simeq
 \frac{1}{3\dlt},
 \qquad
 \Shat(\dlt)
 \simeq
 \frac{1}{81\dlt^2}.
 \label{eq:thin-wall-radius-action}
\end{equation}
These relations will be used as analytic checks on the numerical solutions.

\section{Fluctuation spectrum around the bubble}
\label{sec:fluctuations}

The bounce is a stationary point of the three-dimensional Euclidean action,
but it is not a minimum.  To determine the local geometry of this saddle, we
expand the action about $\phi_b$,
\begin{equation}
 S_3[\phi_b+\eta]
 =
 S_3[\phi_b]
 +\frac{1}{2}
 \int\dd^3x\,
 \eta\,\mathcal{M}_{E}^{\rm phys}\eta
 +O(\eta^3),
 \label{eq:hessian-expansion}
\end{equation}
where the first variation vanishes by the bounce equation and
\begin{equation}
 \mathcal{M}_{E}^{\rm phys}
 =
 -\bm{\nabla}^2
 +V_{\phi\phi}(\phi_b),
 \label{eq:euclidean-hessian}
\end{equation}
is the physical Euclidean Hessian, or quadratic fluctuation operator
\cite{CallanColeman1977,Linde1983}.  Its determinant enters the semiclassical
nucleation prefactor.  Its spectrum also identifies the characteristic
directions around the saddle: one negative $\ell=0$ eigenvalue changes the
bubble radius, the three $\ell=1$ zero modes translate the bubble center, and
positive eigenvalues describe stable Gaussian fluctuations.  The localized
positive branch continuously connected to the planar P\"oschl--Teller shape
mode is the main object of this work.  These eigenvalues are curvatures of the
Euclidean free-energy functional; their relation to real-time post-nucleation
frequencies is discussed separately in \cref{sec:discussion}.  We use the
dimensionless operator and eigenvalues
\begin{equation}
 \widehat{\mathcal M}_E\equiv\Lzero^2\mathcal M_E^{\rm phys},
 \qquad
 \Lam_{\ell n}\equiv\Lzero^2\lambda_{\ell n}^{\rm phys},
 \label{eq:dimensionless-hessian-eigenvalues}
\end{equation}
so every $\Lam_{\ell n}$ quoted below is dimensionless.

Expanding in spherical harmonics and writing
\begin{equation}
 \eta(\rho,\Omega)
 =
 \frac{u_{\ell n}(\rho)}{\rho}Y_{\ell m}(\Omega),
 \end{equation}
we obtain
\begin{equation}
 \left[
 -\frac{\dd^2}{\dd\rho^2}
 +\frac{\ell(\ell+1)}{\rho^2}
 +U(\rho;\dlt)
 \right]u_{\ell n}
 =
 \Lam_{\ell n}u_{\ell n},
 \label{eq:radial-eigenproblem}
\end{equation}
with
\begin{equation}
 U(\rho;\dlt)
 =
 4-24s_b+24s_b^2-12\dlt+24\dlt s_b.
 \label{eq:fluctuation-potential}
\end{equation}
For finite-box diagonalization we impose
$u_{\ell}(0)=u_{\ell}(\rho_{\rm box})=0$; the half-line problem is instead
defined by regularity at the origin and the appropriate asymptotic condition.

The spectral interpretation differs from that of an infinite planar wall.
The spherical radial problem is defined on the half-line
$0\leq\rho<\infty$, and the bounce approaches the false vacuum as
$\rho\to\infty$.  Therefore the continuum threshold of the radial Hessian is
uniquely determined by the false-vacuum asymptotic region,
\begin{equation}
 \Lam_{\rm cont}=\Lam_{\rm f}=4-12\dlt.
 \label{eq:false-continuum-threshold}
\end{equation}
The quantity
\begin{equation}
 \Lam_{\rm t}=4+12\dlt
 \label{eq:true-interior-curvature}
\end{equation}
is instead the local Hessian curvature in an approximately true-vacuum
interior.  Because the interior occupies a finite radial interval, it does not
define a second asymptotic continuum threshold.  An isolated discrete
eigenvalue of the spherical operator satisfies
\begin{equation}
 \Lam_{\ell n}<\Lam_{\rm f}.
 \label{eq:localization-condition}
\end{equation}
At a merger point, however, the $\ell\geq1$ threshold solution behaves as
$u_\ell\propto\rho^{-\ell}$ and remains square integrable on the half-line.
It therefore defines a threshold eigenstate at
$\Lam_{\ell n}=\Lam_{\rm f}$; no 
square-integrable continuation
exists beyond the merger.
In the thin-wall regime, where the interface is locally planar and a broad
true-vacuum core exists, $\Lam_{\rm f}$ and $\Lam_{\rm t}$ nevertheless
determine useful exterior and interior evanescent scales,
\begin{equation}
 \frac{\xi_{\rm f}}{\Lzero}
 =\left(\Lam_{\rm f}-\Lam_{\ell n}\right)^{-1/2},
 \qquad
 \frac{\xi_{\rm t}^{\rm loc}}{\Lzero}
 =\left(\Lam_{\rm t}-\Lam_{\ell n}\right)^{-1/2}.
 \label{eq:localization-lengths}
\end{equation}
The first is the true asymptotic decay length outside the bubble; the second
is a local interior decay scale and becomes progressively less asymptotic as
the bubble enters the thick-wall regime.

The branch continuously connected to the planar translation mode has a simple
thin-wall derivation.  Let the spherical interface be displaced from its
half-height radius according to $R_{1/2}\to R_{1/2}+\zeta(\Omega)$.  At
leading thin-wall order, the stationarity condition for
$F=\sigma A-\Delta V\,\mathcal V$ gives
$\Delta V=2\sigma/R_{1/2}$.  Expanding to quadratic order then yields
\begin{equation}
 \Delta^{(2)}F
 =\frac{\sigma}{2}\int\dd\Omega
 \left[(\bm\nabla_\Omega\zeta)^2-2\zeta^2\right].
 \label{eq:surface-second-variation}
\end{equation}
After division by the wall kinetic normalization, each spherical harmonic
therefore carries
\begin{equation}
 \Lam_{{\rm tr},\ell}
 \simeq
 \frac{\ell(\ell+1)-2}{(R_{1/2}/\Lzero)^2}
 =9\left[\ell(\ell+1)-2\right]\dlt^2.
 \label{eq:translation-thin-wall}
\end{equation}
The $\ell=0$ member is the unique negative mode, the $\ell=1$ member consists
of the translational zero modes, and $\ell\geq2$ describes stable deformations
of the spherical interface.

For a wall-localized shape function centered at
$\rho=R_{1/2}/\Lzero$, the centrifugal term gives
\begin{equation}
 \left\langle\frac{\ell(\ell+1)}{\rho^2}\right\rangle
 =\frac{\ell(\ell+1)}{(R_{1/2}/\Lzero)^2}
 +O\!\left[(R_{1/2}/\Lzero)^{-4}\right].
 \label{eq:centrifugal-expectation}
\end{equation}
The nominal $O[(R_{1/2}/\Lzero)^{-3}]$ term vanishes in the planar limit
because the centered shape-mode probability density is even.  Hence
\begin{equation}
 \Lam_{{\rm sh},\ell}-\Lam_{{\rm sh},0}
 \simeq
 9\ell(\ell+1)\dlt^2.
 \label{eq:shape-curvature-expectation}
\end{equation}

\subsection{Near-coexistence operator expansion}
\label{sec:near-critical-operator}

The absence of an $O(\dlt)$ shift can be understood analytically.  Introduce
the dimensionless half-height radius and wall-centered coordinate
\begin{equation}
 \mathcal R\equiv\frac{R_{1/2}}{\Lzero},
 \qquad
 x=\rho-\mathcal R,
\end{equation}
and expand
\begin{equation}
 \mathcal R^{-1}=3\dlt+O(\dlt^3),
 \qquad
 s_b(\rho)=s_0(x)+\dlt s_1(x)+\dlt^2s_2(x)+\cdots.
 \label{eq:bounce-near-critical-expansion}
\end{equation}
The wall center is fixed by $s_b(\rho=\mathcal R)=1/2$, or equivalently by $s_b(x=0)=1/2$.
The linearization of the bounce equation about the degenerate planar wall is governed by the planar Hessian $\mathcal M_{\rm PT}$ defined in \cref{eq:PT-operator}.  At first order in $\dlt$, the bounce equation reduces to
\begin{equation}
 \mathcal M_{\rm PT}s_1
 =
 6s_0'+12s_0(1-s_0).
 \label{eq:first-order-profile-equation}
\end{equation}
Since $s_0'=-2s_0(1-s_0)$, the right-hand side vanishes.  The only bounded solution compatible with the asymptotic boundary conditions is therefore $s_1=C s_0'$, corresponding to an infinitesimal translation of the planar wall.  Expanding the wall-center condition gives $s_1(0)=0$.  Since $s_0'(0)=-1/2\neq0$, it follows that $C=0$, and hence $s_1=0$.

Denoting by $\widehat{\mathcal M}_{E,\ell}$ the radial restriction of the dimensionless Hessian $\widehat{\mathcal M}_E$ in the $\ell$th partial wave, its wall-centered expansion is
\begin{equation}
 \widehat{\mathcal M}_{E,\ell}
 =\mathcal M_{\rm PT}
 +\dlt\mathcal M_1
 +\dlt^2\left[\mathcal M_2+9\ell(\ell+1)\right]
 +O(\dlt^3),
 \qquad
 \mathcal M_1=-12\tanh x.
 \label{eq:hessian-near-critical-expansion}
\end{equation}
Here $\mathcal M_2$ denotes the $\ell$-independent second-order correction generated by relaxation of the radial background, while the explicit $9\ell(\ell+1)$ term is the leading centrifugal contribution.  Since $\chi_{\rm sh}^2$ is even and $\mathcal M_1$ is odd,
\begin{equation}
 \Lam_{\rm sh}^{(1)}
 =\langle\chi_{\rm sh}|\mathcal M_1|\chi_{\rm sh}\rangle=0.
 \label{eq:first-order-shape-zero}
\end{equation}
The centrifugal term starts at $\mathcal R^{-2}=9\dlt^2+O(\dlt^4)$, so all
angular branches lack a linear correction.  This absence of a first-order
eigenvalue shift was already obtained in the perturbative thin-wall analysis
of M\"unster and Rotsch \cite{MuensterRotsch2000};
\cref{eq:first-order-profile-equation,eq:first-order-shape-zero} provide a
short rederivation adapted to the centered parametrization used here.

The common quadratic coefficient may be written in standard second-order
perturbation theory as
\begin{equation}
 C_{\rm bg}
 =\langle\chi_{\rm sh}|\mathcal M_2|\chi_{\rm sh}\rangle
 +\sum_{a\neq {\rm sh}}
 \frac{|\langle a|\mathcal M_1|\chi_{\rm sh}\rangle|^2}
 {3-\Lam_a},
 \label{eq:second-order-shape-decomposition}
\end{equation}
where the sum includes the translational state and the continuum.
For the present quartic family this expression is known analytically.
Using the exact convention map collected in \cref{app:MR-map}, the two
discrete bands in the M\"unster--Rotsch conventions are
\begin{align}
 \omega_{0\ell}
 &=\widetilde\eta^2[\ell(\ell+1)-2]+O(\widetilde\eta^4),
 \\
 \omega_{3\ell}
 &=3+\widetilde\eta^2[\ell(\ell+1)+3-\pi^2]
 +O(\widetilde\eta^4).
 \label{eq:MR-discrete-spectrum}
\end{align}
Using \cref{eq:MR-spectral-map} gives
\begin{align}
 \Lam_{{\rm tr},\ell}
 &=9[\ell(\ell+1)-2]\dlt^2+O(\dlt^4),
 \label{eq:analytic-translation-spectrum}
 \\
 \Lam_{{\rm sh},\ell}
 &=3+9[\ell(\ell+1)+4-\pi^2]\dlt^2+O(\dlt^4).
 \label{eq:analytic-shape-spectrum}
\end{align}
Consequently,
\begin{equation}
 C_{\rm bg}=9(4-\pi^2)=-52.8264396\ldots,
 \label{eq:analytic-common-coefficient}
\end{equation}
while $9\ell(\ell+1)$ is the geometrical angular splitting.  The numerical
calculation below tests these analytic coefficients and then follows the
individual eigenstates beyond the domain controlled by the thin-wall series.

\section{Numerical method}
\label{sec:numerics}

\subsection{Bounce solver}

Equation~\eqref{eq:bounce-equation} is solved as a singular boundary-value problem using an adaptive fourth-order collocation method.  The radial regularity condition is imposed analytically at $\rho=0$.  A thin-wall profile centered at
$\rho=1/(3\dlt)$ is used as the initial iterate.  The outer boundary is chosen as
\begin{equation}
 \rho_{\rm BVP}
 =
 \max\left(25,\frac{1}{3\dlt}+20\right),
 \label{eq:domain-choice}
\end{equation}
the residual tolerance is $2\times10^{-9}$, and the adaptive mesh is allowed up to $10^5$ nodes.  The scan covers
\begin{equation}
 \dlt=
 0.005,\,0.0075,\,0.01,\,0.015,\,0.02,\,0.03,\,0.05,
 \,0.075,\,0.10,\,0.125,\,0.15,\,0.18,\,0.20.
 \label{eq:scan-points}
\end{equation}
The action in \cref{eq:dimensionless-action} is evaluated by adaptive quadrature.

\subsection{Radial eigenvalue solver}

The operator in \cref{eq:radial-eigenproblem} is discretized on
$0\leq\rho\leq\rho_{\rm box}$ at grid points $\rho_j=jh$,
$j=0,\ldots,N$, with $Nh=\rho_{\rm box}$, using a centered second-order
finite difference.  Unless varied in the convergence tests,
$\rho_{\rm box}=\rho_{\rm BVP}$.
For the reduced radial function, regularity of the original fluctuation at the
origin requires
\begin{equation}
 u_{\ell n}(\rho)\propto\rho^{\ell+1}
 \qquad (\rho\to0).
 \label{eq:origin-regularity}
\end{equation}
Numerically, the endpoint value $u_0=0$ is eliminated from the matrix, the
centrifugal term is evaluated only at the interior points $\rho_j>0$, and
$u_N=0$ is imposed at the outer boundary.  This implementation selects the
regular solution in \cref{eq:origin-regularity} and avoids evaluating
$\ell(\ell+1)/\rho^2$ at the origin.  
The resulting real symmetric tridiagonal matrix is diagonalized using a standard tridiagonal eigensolver.

The low-lying eigenvalues for $\ell=0,1,2$ are computed at grid spacings
$h=0.01$ and $h=0.005$ and combined by Richardson extrapolation,
\begin{equation}
 \Lam_{\rm ext}
 =
 \frac{4\Lam(h/2)-\Lam(h)}{3}.
 \label{eq:richardson}
\end{equation}
The $\ell=1$ translational zero mode is an internal accuracy test.  Its
extrapolated magnitude is below $3\times10^{-9}$ for every point in the scan.
Independent calculations with $\rho_{\rm box}$ increased by $5$ and $10$ change
the extrapolated $\ell=0$ shape eigenvalue by at most
$9.4\times10^{-10}$ at the tested points $\dlt=0.05$, $0.10$, and $0.20$.

The radial eigenfunctions are normalized by
$\int_0^{\rho_{\rm box}}|u_{\ell n}(\rho)|^2\dd\rho=1$.  The shape branch is
identified by eigenvalue continuity, node structure, and the overlap
\begin{equation}
 \mathcal O_{\rm sh}(\dlt)
 =\left|\int_0^{\rho_{\rm box}}\dd\rho\,
 u_{{\rm sh},0}(\rho;\dlt)
 \chi_{\rm sh}(\rho-R_{1/2}/\Lzero)\right|,
 \label{eq:shape-overlap-definition}
\end{equation}
where the shifted planar function is renormalized on the numerical interval.
The sign of $u_{{\rm sh},0}$ is chosen so that the unmodified overlap is
positive.  The displacement of the radial probability density is measured by
\begin{equation}
 \langle x\rangle
 =\int_0^{\rho_{\rm box}}\dd\rho\,
 (\rho-R_{1/2}/\Lzero)|u_{{\rm sh},0}(\rho)|^2.
 \label{eq:centroid-definition}
\end{equation}
Because $\eta=uY_{\ell m}/\rho$ and
$\int|Y_{\ell m}|^2\dd\Omega=1$, this is the radial measure induced by the
three-dimensional fluctuation norm in dimensionless variables.

\subsection{Threshold shooting and continuum-merger criterion}
\label{sec:threshold-shooting}

The Dirichlet diagonalization is reliable for an isolated eigenvalue below the
continuum, but an above-threshold box level is not a discrete eigenvalue of the
half-line operator.  We therefore determine the threshold endpoints of the
$\ell=1,2$ positive branches directly at the common threshold
$\Lam=\Lam_{\rm f}$.  Writing
\begin{equation}
 u_\ell(\rho)=\rho^{\ell+1}w_\ell(\rho)
\end{equation}
in the threshold radial equation gives
\begin{equation}
 w_\ell''+\frac{2(\ell+1)}{\rho}w_\ell'
 =24s_b(\rho)\bigl[s_b(\rho)-1+\dlt\bigr]w_\ell.
 \label{eq:threshold-w-equation}
\end{equation}
Outside the bubble, where the right-hand side vanishes,
\begin{equation}
 w_\ell(\rho)=A_\ell+B_\ell\rho^{-(2\ell+1)}.
\end{equation}
The coefficient of the growing threshold solution can therefore be extracted
at a finite matching radius as
\begin{equation}
 A_\ell(\rho_{\rm m})
 =w_\ell(\rho_{\rm m})
 +\frac{\rho_{\rm m}}{2\ell+1}w_\ell'(\rho_{\rm m}).
 \label{eq:threshold-A}
\end{equation}
The bound-state branch reaches its threshold endpoint when
\begin{equation}
 A_\ell(\dlt_{\rm merge})=0.
 \label{eq:merge-condition}
\end{equation}
At this zero, the asymptotic radial function is
\begin{equation}
 u_\ell(\rho)=\rho^{\ell+1}w_\ell(\rho)
 \sim B_\ell\rho^{-\ell}.
 \label{eq:threshold-eigenstate-tail}
\end{equation}
For $\ell=1,2$ this tail belongs to $L^2(0,\infty)$, so the endpoint is a
threshold eigenstate.  On the post-merger side, $A_\ell\neq0$ restores the
growing term $u_\ell\sim A_\ell\rho^{\ell+1}$ and there is no normalizable
half-line continuation.  The regular origin expansion is used to initialize
the integration.  The
threshold solution has one radial node in both sectors, identifying it with
the band connected to the planar shape state rather than the lower
displacement band.

For the reference calculation, the bounce is resolved on
$0\leq\rho\leq\rho_{\rm BVP}=70$ with collocation tolerance $5\times10^{-10}$, and
\cref{eq:threshold-w-equation} is integrated with an adaptive eighth-order
Runge--Kutta method using relative and absolute tolerances
$2\times10^{-12}$ and $2\times10^{-14}$.  The zero of $A_\ell$ is bracketed
on a 13-point interval of width $0.006$ and located by Brent interpolation.
Independent fixed-step fourth-order Runge--Kutta calculations vary the step
from $h=0.04$ to $0.00125$ and the matching radius $\rho_{\rm m}$ from $12$ to $60$.
Finite-box diagonalizations on both sides of each root provide a separate
localization check.

\subsection{Near-coexistence fitting procedure}
\label{sec:fit-procedure}

The asymptotic coefficients quoted below are obtained by unweighted linear
least squares after subtracting the known leading thin-wall or planar term.
The central fit uses the six points
\begin{equation}
 \dlt=0.005,\ 0.0075,\ 0.01,\ 0.015,\ 0.02,\ 0.03.
\end{equation}
We use the bases $(\dlt,\dlt^3)$ for
$R_{1/2}/\Lzero-1/(3\dlt)$, $(1,\dlt^2)$ for
$\Shat-1/(81\dlt^2)$, and $(\dlt^2,\dlt^4)$ for the wall width and eigenvalue
shifts.  Because the data are deterministic numerical solutions rather than
repeated measurements, we use the maximum fit residual and the variation of
the leading coefficient under changes of the upper fit limit as convergence
diagnostics rather than formal statistical confidence intervals.  The
corresponding values are collected in \cref{tab:fit-diagnostics}.

\subsection{Geometric wall-crossover diagnostic}
\label{sec:wall-crossover-method}

To separate spectral loss of a localized state from geometric loss of a
recognizable bubble wall, we continue the reduced-model bounce beyond the
main scan until the bubble center reaches the half-height value,
\begin{equation}
 s_b(0)=\frac12.
 \label{eq:wall-crossover-criterion}
\end{equation}
At this point a positive half-height radius ceases to exist, so the
wall-centered coordinate used for the localized-mode diagnostics is no longer
well defined.  The same collocation and Hessian solvers are used throughout
this continuation.  This criterion is formulated directly in the reduced
quartic variables and does not require assigning a benchmark temperature.

\section{Results}
\label{sec:results}

\subsection{Bounce profiles and geometric observables}

\begin{figure}[t]
 \centering
 \begin{minipage}[t]{0.49\textwidth}
  \centering
  \includegraphics[width=\linewidth]{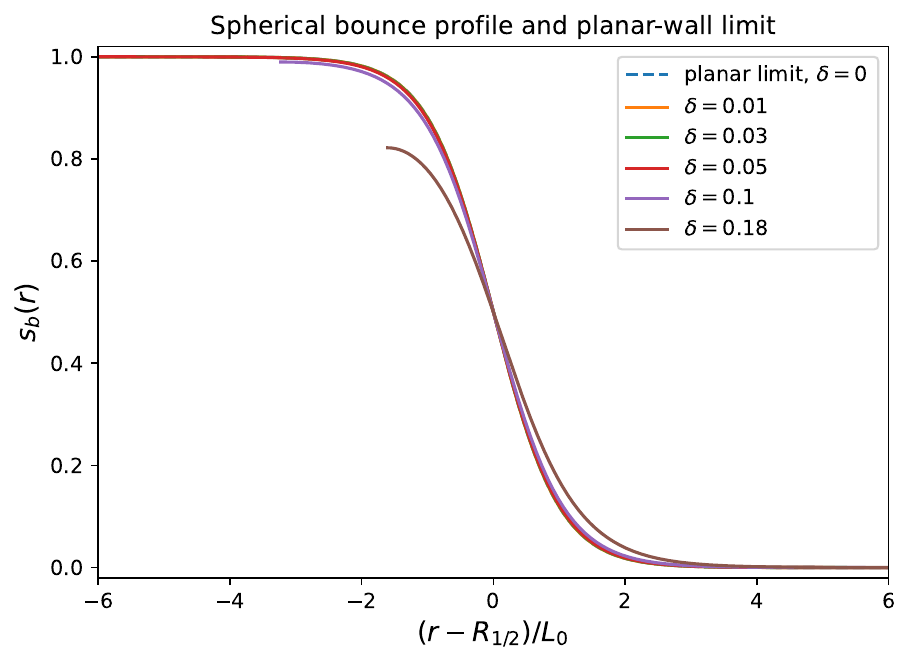}\\[-1mm]
  \small (a) Bounce profiles centered at $R_{1/2}$.
 \end{minipage}
 \hfill
 \begin{minipage}[t]{0.49\textwidth}
  \centering
  \includegraphics[width=\linewidth]{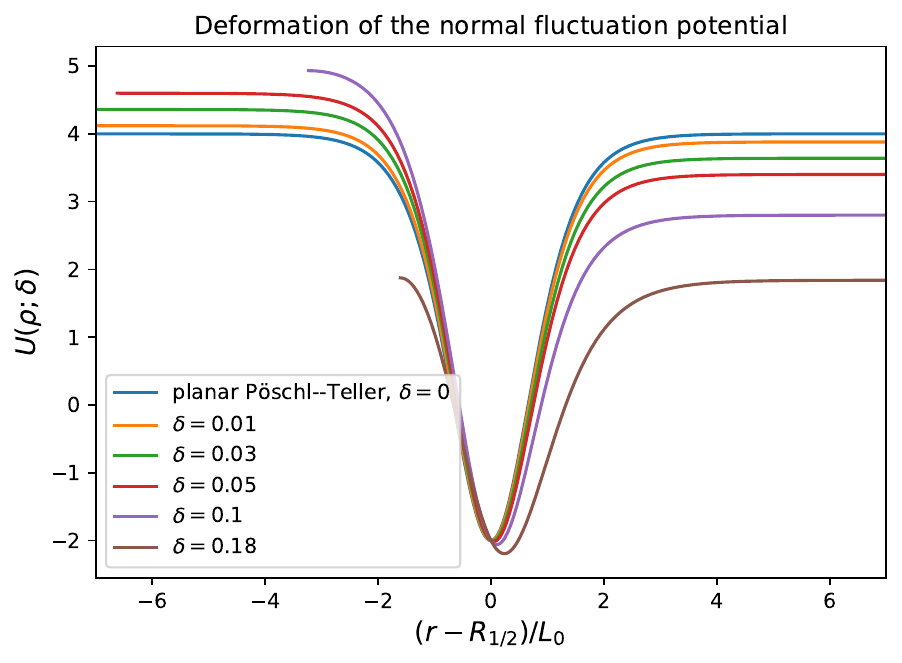}\\[-1mm]
  \small (b) Normal fluctuation potentials.
 \end{minipage}
 \caption{Deformation away from the exact planar wall.  The dotted curves denote the $\dlt=0$ P\"oschl--Teller limit, and the solid curves show spherical critical bubbles at finite $\dlt$.}
 \label{fig:profiles-potentials}
\end{figure}

\Cref{fig:profiles-potentials} compares the spherical bounce profiles and fluctuation potentials with their planar limits.  Near $\dlt=0$, the wall is locally indistinguishable from the exact kink.  Increasing $\dlt$ reduces the radius, broadens the wall, and makes the fluctuation potential increasingly asymmetric between its true- and false-vacuum limits.

\begin{figure}[t]
 \centering
 \begin{minipage}[t]{0.49\textwidth}
  \centering
  \includegraphics[width=\linewidth]{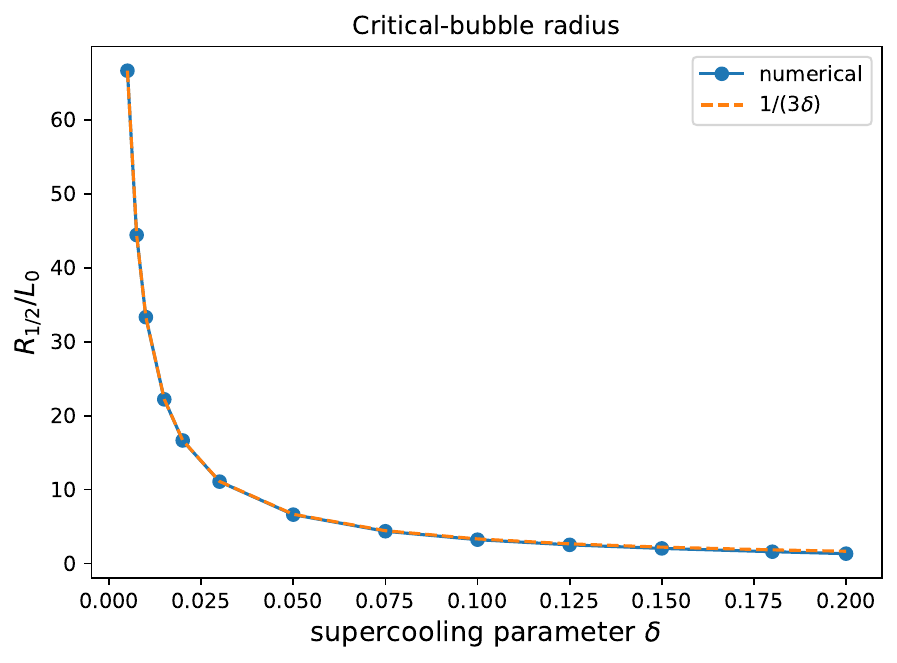}\\[-1mm]
  \small (a) Bubble radius.
 \end{minipage}
 \hfill
 \begin{minipage}[t]{0.49\textwidth}
  \centering
  \includegraphics[width=\linewidth]{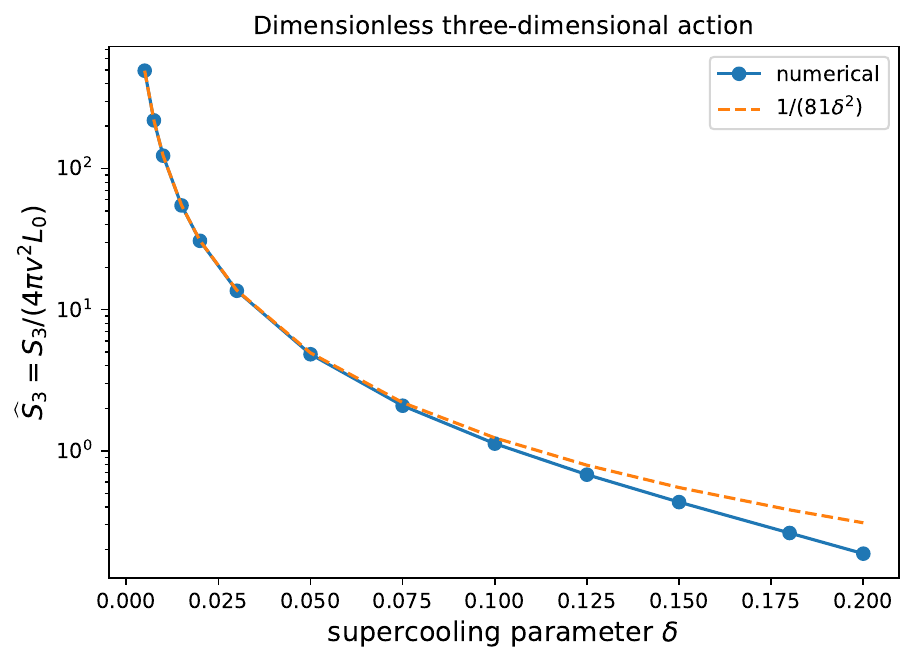}\\[-1mm]
  \small (b) Dimensionless three-dimensional action $\Shat$.
 \end{minipage}
 \caption{Numerical results compared with the thin-wall limits in \cref{eq:thin-wall-radius-action}.}
 \label{fig:radius-action}
\end{figure}

The bubble radius and dimensionless three-dimensional action are shown in \cref{fig:radius-action}.  Fits restricted to $\dlt\leq0.03$ yield
\begin{align}
 \frac{R_{1/2}}{\Lzero}
 &=
 \frac{1}{3\dlt}
 -0.9674\,\dlt
 -5.84\,\dlt^3
 +O(\dlt^5),
 \label{eq:radius-fit}
 \\
 \Shat(\dlt)
 &=
 \frac{1}{81\dlt^2}
 -0.10749
 -0.324\,\dlt^2
 +O(\dlt^4).
 \label{eq:action-fit}
\end{align}
The agreement with the leading thin-wall expressions is excellent in the
near-coexistence regime.  Complete residual and fit-window diagnostics are
listed in \cref{tab:fit-diagnostics}.

\begin{figure}[t]
 \centering
 \includegraphics[width=0.66\textwidth]{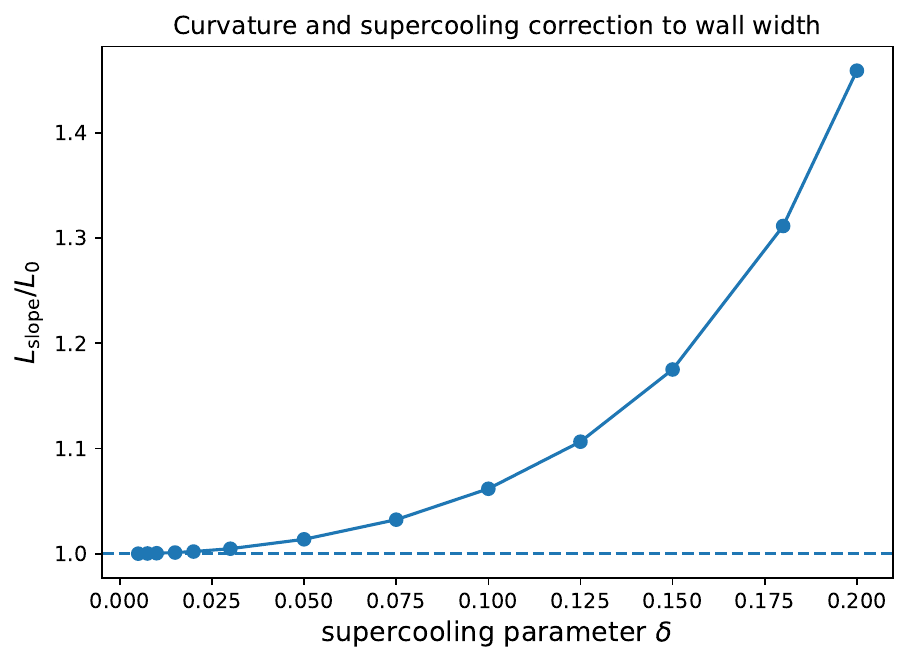}
 \caption{Slope-defined wall width as a function of the supercooling parameter.  The horizontal line is the exact planar value $L_{\rm slope}=\Lzero$.}
 \label{fig:wall-width}
\end{figure}

The slope width changes more slowly than the radius.
The slope-defined wall width is shown in \cref{fig:wall-width}.
Its near-coexistence fit is
\begin{equation}
 \frac{L_{\rm slope}}{\Lzero}
 =
 1+5.317\,\dlt^2
 +74.0\,\dlt^4
 +O(\dlt^6).
 \label{eq:width-fit}
\end{equation}
The absence of a linear term is not merely numerical: the first-order profile correction vanishes after the wall center is fixed, as shown in \cref{sec:near-critical-operator}.  The leading change in the local wall thickness is therefore quadratic even though the vacuum-energy difference is linear in $\dlt$.

For later interpretation, we distinguish three regimes.  The asymptotic
thin-wall fits use $\dlt\leq0.03$, where
$R_{1/2}/L_{\rm slope}\gtrsim11$.  We call
$0.03<\dlt\leq0.18$ the finite-radius wall-resolved regime; over this
range the half-height radius exists and
$R_{1/2}/L_{\rm slope}>1.2$.  The point $\dlt=0.20$, for which this ratio has
fallen below unity, is retained as the endpoint of the main numerical scan rather
than as a thin-wall configuration.  The continuation above $0.20$ to the
central-field-defined geometric crossover, discussed in
\cref{sec:wall-crossover}, is diagnostic only.

\subsection{Displacement branch and universal curvature splitting}

\begin{figure}[t]
 \centering
 \includegraphics[width=0.72\textwidth]{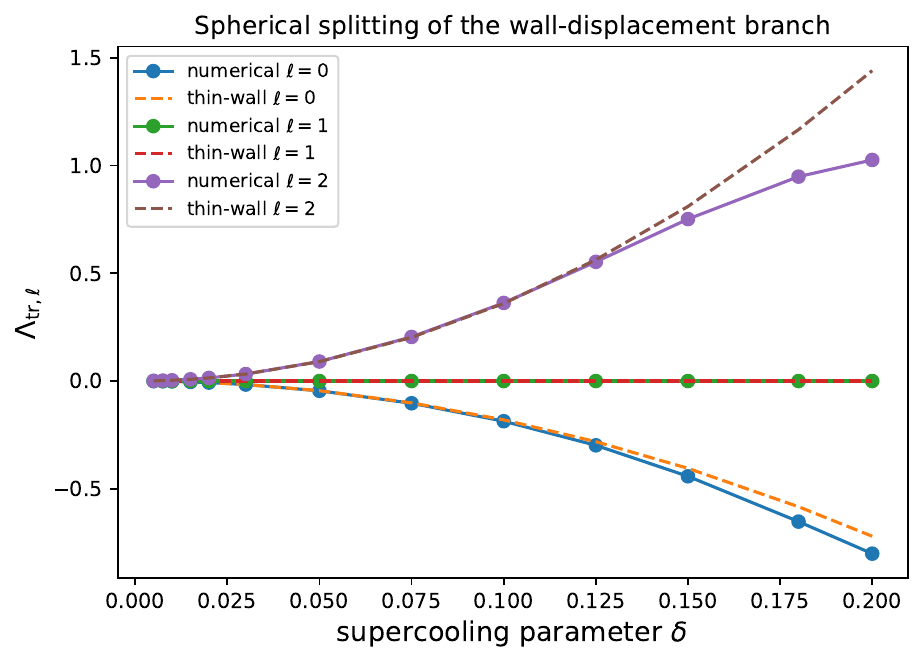}
 \caption{Spherical splitting of the branch connected to the planar translational mode.  Dashed curves show the leading thin-wall formula $9[\ell(\ell+1)-2]\dlt^2$.}
 \label{fig:translation-branch}
\end{figure}

\Cref{fig:translation-branch} shows the lowest eigenvalue in the sectors $\ell=0,1,2$.  The numerical results reproduce the expected structure: one negative mode for $\ell=0$, an exact translational zero mode for $\ell=1$, and a positive quadrupolar deformation for $\ell=2$.  Near coexistence, the numerical behavior may be written as
\begin{align}
 \Lam_{{\rm tr},0}
 &=-18\,\dlt^2-53.4\,\dlt^4+O(\dlt^6),
 \label{eq:translation-l0-fit}
 \\
 \Lam_{{\rm tr},2}
 &=36\,\dlt^2+97.6\,\dlt^4+O(\dlt^6),
 \label{eq:translation-l2-fit}
\end{align}
where the leading coefficients have been set to their analytic thin-wall
values.  The corresponding unconstrained fits give $-17.9997$ and $36.0016$,
respectively, consistent with \cref{eq:translation-thin-wall}; the rounded
coefficients and fit-window variations are reported in
\cref{tab:fit-diagnostics}.

We denote the negative radial eigenvalue by
$\Lam_-\equiv\Lam_{{\rm tr},0}$.  It also satisfies the geometrical relation
\begin{equation}
 \Lam_-
 \simeq
 -\frac{2}{(R_{1/2}/\Lzero)^2}
 \label{eq:negative-mode-radius}
\end{equation}
in the thin-wall regime.  For example, at $\dlt=0.05$, the right-hand side gives $-0.04567$, compared with the numerical result $-0.04534$.

\subsection{Near-coexistence deformation of the shape band}
\label{sec:shape-near-critical}

\begin{figure}[t]
 \centering
 \includegraphics[width=0.68\textwidth]{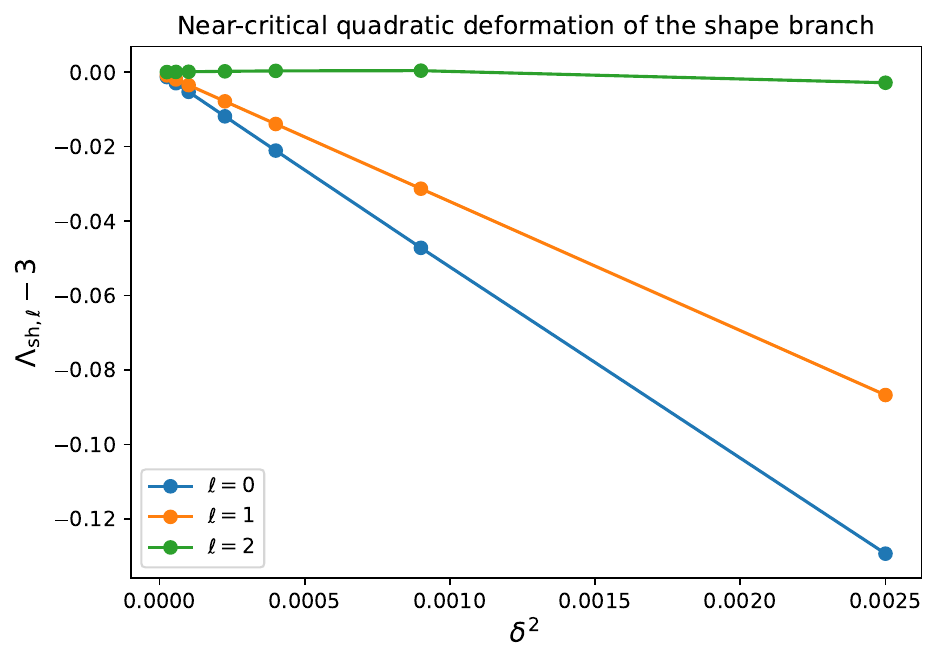}
 \caption{Near-coexistence quadratic deformation of the band connected to the planar P\"oschl--Teller shape mode.}
 \label{fig:shape-near-critical}
\end{figure}

At $\dlt=0$, the three angular sectors become degenerate at the planar value
$\Lam=3$.  At finite $\dlt$, the $\ell=0$ radial operator changes because
the bounce radius, wall core, and vacuum curvatures change together.  We call
the resulting common contribution the radial-background shift.  The explicit
centrifugal term then produces the additional angular splitting.  The
near-coexistence scaling is shown in \cref{fig:shape-near-critical}.

Fits over $\dlt\leq0.03$ give
\begin{align}
 \Lam_{{\rm sh},0}
 &=3+9(4-\pi^2)\dlt^2+451\,\dlt^4+O(\dlt^6),
 \label{eq:shape-l0-fit}
 \\
 \Lam_{{\rm sh},1}
 &=3+9(6-\pi^2)\dlt^2+58.9\,\dlt^4+O(\dlt^6).
 \label{eq:shape-l1-fit}
\end{align}
Here the leading coefficients have been fixed to the analytic values in
\cref{eq:analytic-shape-spectrum}; unconstrained fits give $-52.82$ and
$-34.826$, respectively.  The branch splittings likewise satisfy
\begin{align}
 \Lam_{{\rm sh},1}-\Lam_{{\rm sh},0}
 &=18\,\dlt^2+O(\dlt^4),
 \\
 \Lam_{{\rm sh},2}-\Lam_{{\rm sh},0}
 &=54\,\dlt^2+O(\dlt^4),
\end{align}
with unconstrained numerical coefficients $17.998$ and $54.018$.  Thus the
leading result is the known analytic expression
\begin{equation}
 \Lam_{{\rm sh},\ell}
 =3+9[\ell(\ell+1)+4-\pi^2]\dlt^2+O(\dlt^4).
 \label{eq:shape-master-fit}
\end{equation}
This agreement establishes the controlled near-coexistence anchor from which we follow the individual states to their partial-wave-dependent continuum thresholds.

\subsection{Angular-momentum-dependent continuum mergers}
\label{sec:shape-mergers}

\begin{figure}[t]
\centering
\begin{minipage}{0.49\textwidth}
  \centering
  \includegraphics[width=\linewidth]{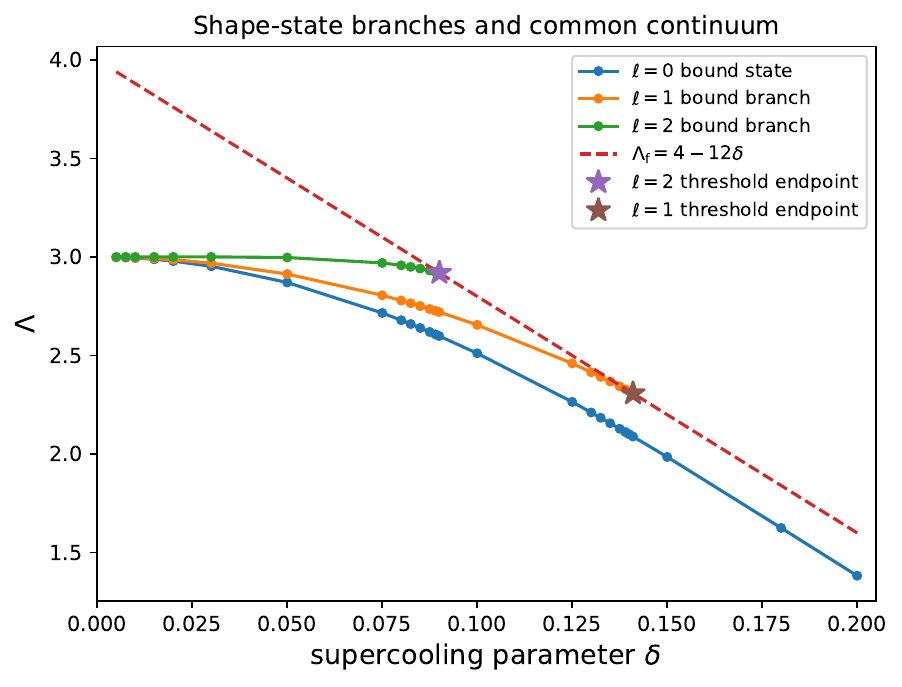}

  \small (a) Shape-state branches and common continuum.
\end{minipage}
\hfill
\begin{minipage}{0.49\textwidth}
  \centering
  \includegraphics[width=\linewidth]{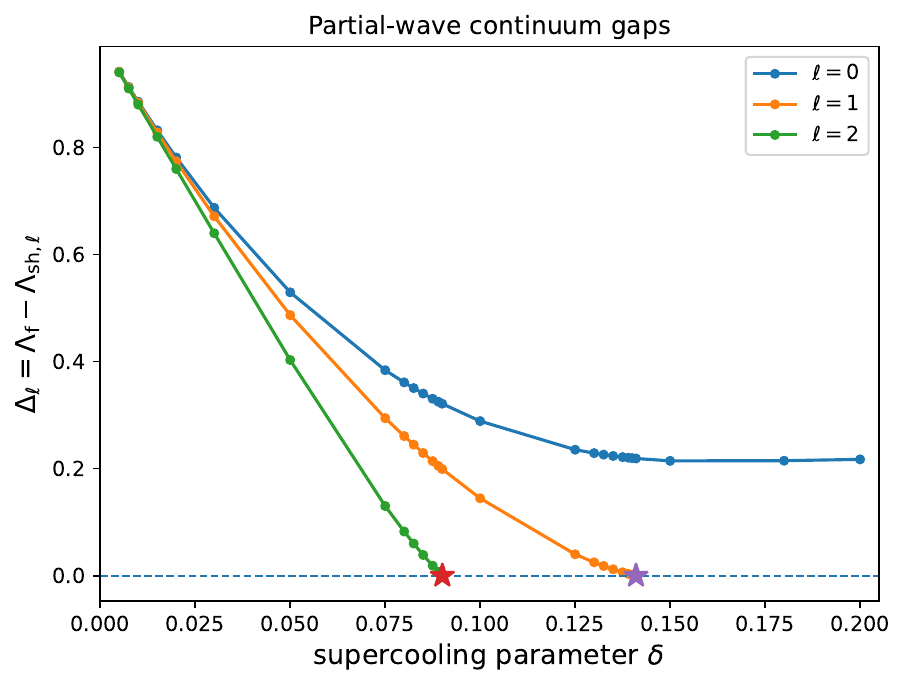}

  \small (b) Partial-wave continuum gaps.
\end{minipage}

\caption{
Angular-momentum-dependent continuation of the spectral band connected
to the planar shape mode.
(a) Bound-state branches and their threshold endpoints, together with the false-vacuum continuum threshold, which is common to all partial waves.
(b) Corresponding continuum gaps.
The $\ell=2$ and $\ell=1$ curves terminate at their square-integrable
threshold eigenstates determined by shooting.
Finite-box levels above the threshold are not plotted as discrete eigenstates.
}
\label{fig:shape-mergers}
\end{figure}

The continuum threshold is $\Lam_{\rm f}=4-12\dlt$ for every partial wave.
\Cref{fig:shape-mergers}(a) shows the partial-wave-resolved shape-state branches and the common false-vacuum continuum threshold, while \cref{fig:shape-mergers}(b) displays the corresponding continuum gaps.
The
threshold-shooting values and their conservative numerical uncertainties are
listed in \cref{tab:shape-merger-summary}.  The $\ell=2$ state reaches the
continuum first because of the larger centrifugal barrier, while the $\ell=0$ shape
state remains below threshold throughout the computed reduced-model scan.

%

The leading thin-wall equation obtained by equating
\cref{eq:analytic-shape-spectrum} to $\Lam_{\rm f}$ predicts
\begin{equation}
 \dlt_{{\rm merge},2}^{\rm TW}=0.0826650,
 \qquad
 \dlt_{{\rm merge},1}^{\rm TW}=0.1411769.
\end{equation}
The $\ell=1$ estimate is already accurate to $1.6\times10^{-4}$, whereas
the full $\ell=2$ merger is shifted upward by $7.38\times10^{-3}$, showing
that finite-supercooling corrections are important for this sector.  This
comparatively larger correction is consistent with the near cancellation in
the quadratic coefficient $9(10-\pi^2)$ of the $\ell=2$ branch.

\begin{table}[t]
\centering
\caption{Continuum-merger positions of the positive angular shape states.  At each endpoint the square-integrable threshold eigenstate has one radial node.}
\label{tab:shape-merger-summary}
\begin{tabular}{ccccc}
\hline
$\ell$ & $\dlt_{\rm merge}$ & numerical uncertainty & thin-wall estimate & full minus thin wall \\
\hline
2 & 0.0900472 & $2\times10^{-8}$ & 0.0826650 & $+0.0073821$ \\
1 & 0.1410162 & $2\times10^{-8}$ & 0.1411769 & $-0.0001607$ \\
\hline
\end{tabular}
\end{table}

Finite-box calculations on both sides of each endpoint independently
confirm this interpretation; numerical details are given in
\cref{app:merger-convergence}.

\subsection{Continuum threshold, interior curvature, and asymmetric localization}

\begin{figure}[t]
 \centering
 \includegraphics[width=0.72\textwidth]{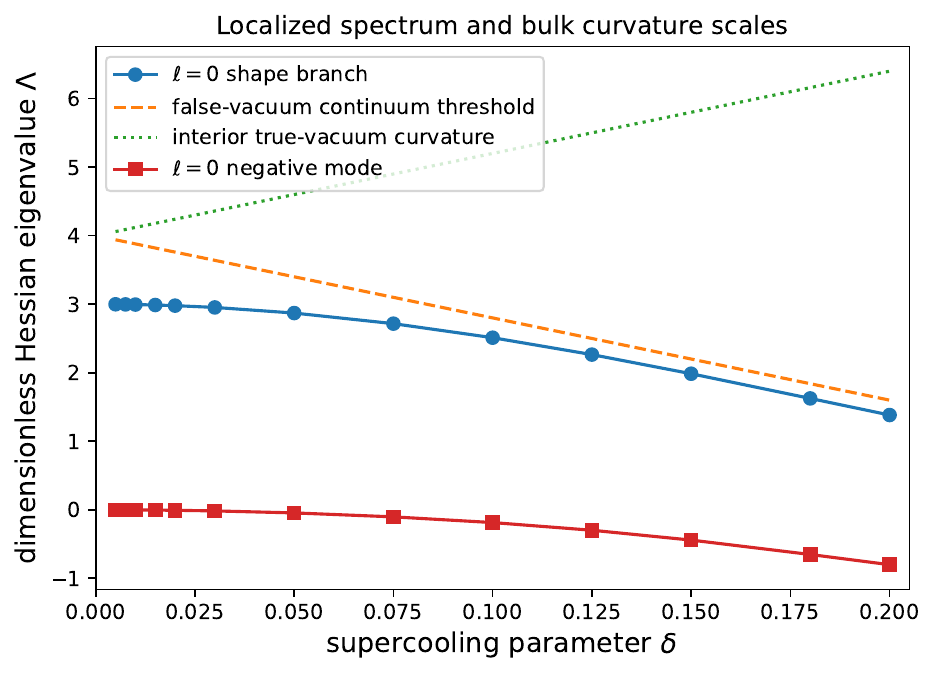}
 \caption{The $\ell=0$ localized branches, the false-vacuum continuum threshold $\Lam_{\rm f}$, and the interior true-vacuum curvature $\Lam_{\rm t}$.  Only $\Lam_{\rm f}$ is an asymptotic continuum threshold of the spherical radial operator.}
 \label{fig:eigen-threshold}
\end{figure}

\Cref{fig:eigen-threshold} compares the $\ell=0$ shape eigenvalue with the false-vacuum continuum threshold and the local true-vacuum curvature in the bubble interior.  Only $\Lam_{\rm f}$ is a continuum threshold of the spherical half-line operator.  It decreases linearly with $\dlt$, but the shape mode remains below it throughout the computed interval.  The smallest gap occurs near $\dlt\simeq0.15$ and remains positive,
\begin{equation}
 \min_{0.005\leq\dlt\leq0.20}
 \left(\Lam_{\rm f}-\Lam_{{\rm sh},0}\right)
 \simeq0.2145.
 \label{eq:minimum-gap}
\end{equation}
Thus the $\ell=0$ shape state does not merge into the continuum before $\dlt=0.20$ in this model.  This statement does not apply to the $\ell=1,2$ angular states, whose mergers were determined in \cref{sec:shape-mergers}.

The spatial asymmetry has a direct bulk-mass interpretation.  Far outside the bubble, where $s_b\to0$, the Hessian approaches the false-vacuum curvature $\Lam_{\rm f}=4-12\dlt$.  In a sufficiently broad true-vacuum core, where $s_b\simeq1$, it approaches the larger local interior curvature $\Lam_{\rm t}=4+12\dlt$.  For the radial shape state, which we denote by
$\Lam_{\rm sh}\equiv\Lam_{{\rm sh},0}$ in this subsection, the exterior asymptotic tail and the locally planar interior tail have the form
\begin{equation}
 u_{\rm sh}(x)
 \propto
 \begin{cases}
  \exp(-\kappa_{\rm f}x), & x>0 \quad \text{(false-vacuum exterior)},\\[2mm]
  \exp(+\kappa_{\rm t}x), & x<0 \quad \text{(true-vacuum interior)},
 \end{cases}
 \qquad
 \kappa_{\rm f,t}
 =\sqrt{\Lam_{\rm f,t}-\Lam_{\rm sh}},
 \label{eq:asymmetric-tails}
\end{equation}
where $x=(r-R_{1/2})/\Lzero$.  Since $\Lam_{\rm f}<\Lam_{\rm t}$ for every $\dlt>0$, one necessarily has
\begin{equation}
 \kappa_{\rm f}<\kappa_{\rm t},
 \qquad
 \frac{\xi_{\rm f}}{\Lzero}=\kappa_{\rm f}^{-1}
 >
 \frac{\xi_{\rm t}^{\rm loc}}{\Lzero}=\kappa_{\rm t}^{-1}.
 \label{eq:tail-hierarchy}
\end{equation}
Supercooling therefore softens bulk fluctuations in the metastable phase and stiffens them in the stable phase.  The evanescent shape-mode wave function can penetrate more deeply into the lower-mass false-vacuum exterior, whereas it is more strongly suppressed in the higher-mass true-vacuum interior.  The asymmetric leakage is thus not an incidental feature of the numerical eigenfunction, but the expected consequence of the unequal curvatures of the two vacua.  Spherical curvature and the deformation of the wall core alter the eigenvalue and the central profile, while the ordering of the two exponential tail lengths follows directly from the bulk mass hierarchy.  For the largest values of $\dlt$, where the bubble radius is only a few wall widths, finite-radius effects also influence the interior tail; nevertheless, the same hierarchy remains visible in the numerical solutions.

\begin{figure}[t]
 \centering
 \begin{minipage}[t]{0.49\textwidth}
  \centering
  \includegraphics[width=\linewidth]{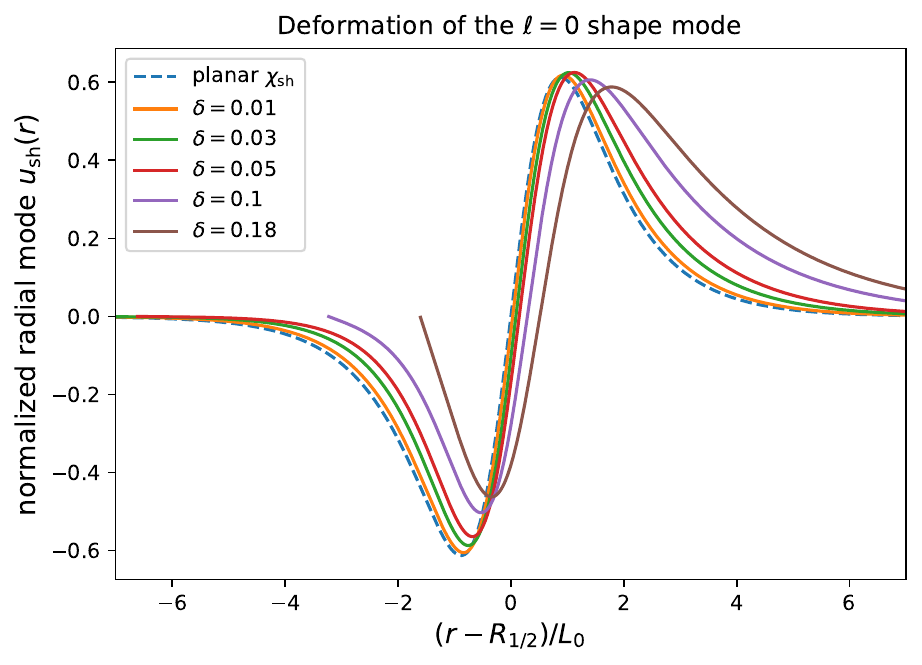}\\[-1mm]
  \small (a) Normalized radial functions $u_{{\rm sh},0}$.
 \end{minipage}
 \hfill
 \begin{minipage}[t]{0.49\textwidth}
  \centering
  \includegraphics[width=\linewidth]{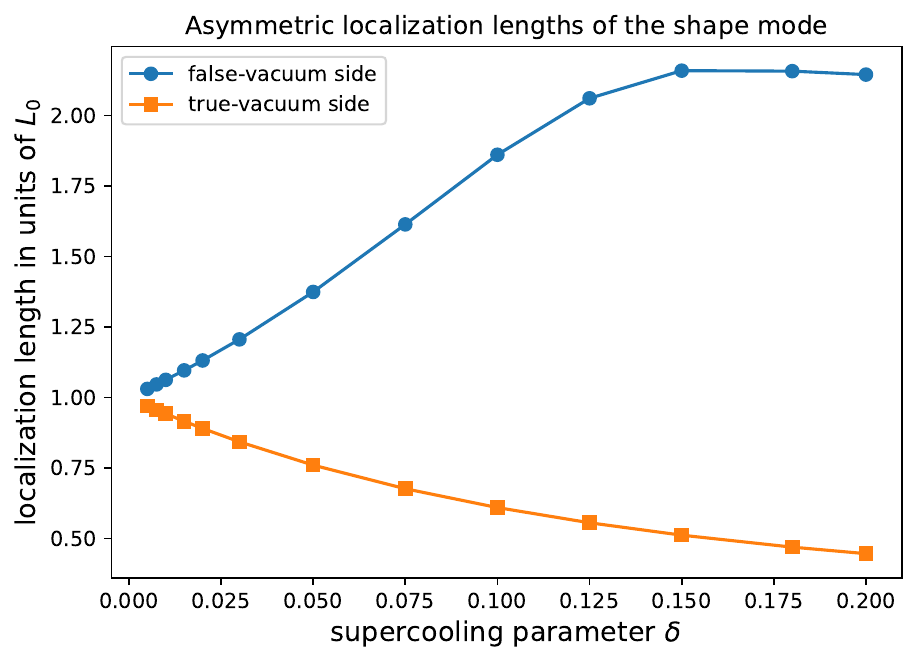}\\[-1mm]
  \small (b) Exterior asymptotic and interior local decay lengths.
 \end{minipage}
 \caption{Spatial deformation and asymmetric localization of the shape mode.  The plotted eigenfunction is the reduced radial function $u$ in $\eta=uY_{\ell m}/\rho$, normalized by $\int|u|^2\dd\rho=1$.}
 \label{fig:mode-localization}
\end{figure}

\Cref{fig:mode-localization} displays this behavior explicitly.  The false-vacuum tail becomes longer, while the true-vacuum tail becomes shorter.  Correspondingly, the overlap with the planar shape function decreases from $0.9981$ at $\dlt=0.01$ to $0.6821$ at $\dlt=0.18$.  The mean location of the radial probability density relative to the half-height radius shifts from $0.129\Lzero$ to $1.881\Lzero$ over the same interval.  The mode is therefore displaced toward the bubble exterior while its eigenvalue remains below the false-vacuum continuum threshold.  Here the reported centroid is the quantity in \cref{eq:centroid-definition}.

\Cref{tab:representative-results} lists selected points from the scan.  The $\ell=1$ zero mode remains numerically consistent with zero, while the $\ell=0$ shape state stays below $\Lam_{\rm f}$.

\begin{table}[t]
\centering
\caption{Representative numerical results.  Lengths are quoted relative to
$\Lzero$.  The $\Lam$ values are dimensionless; the corresponding physical
Hessian eigenvalues are $\Lam/\Lzero^2$.  The ratio
$R_{1/2}/L_{\rm slope}$ is included as a diagnostic of whether a distinct
wall and interior core can be identified.}
\label{tab:representative-results}
\resizebox{\textwidth}{!}{%
\begin{tabular}{ccccccccc}
\hline
$\dlt$ & $R_{1/2}/\Lzero$ & $L_{\rm slope}/\Lzero$ & $R_{1/2}/L_{\rm slope}$ & $\Shat$ & $\Lam_-$ & $\Lam_{{\rm sh},0}$ & $\Lam_{{\rm tr},1}$ & $\Lam_{\rm f}$ \\
\hline
0.010 & 33.32365 & 1.00053 & 33.306 & 123.34927 & -0.00180 & 2.99472 & $<3\times10^{-9}$ & 3.880 \\
0.030 & 11.08193 & 1.00485 & 11.028 &  13.60964 & -0.01624 & 2.95282 & $<3\times10^{-9}$ & 3.640 \\
0.050 &  6.61756 & 1.01377 &  6.528 &   4.82996 & -0.04534 & 2.87069 & $<3\times10^{-9}$ & 3.400 \\
0.100 &  3.23020 & 1.06181 &  3.042 &   1.12354 & -0.18635 & 2.51128 & $<3\times10^{-9}$ & 2.800 \\
0.150 &  2.04779 & 1.17513 &  1.743 &   0.43263 & -0.44190 & 1.98552 & $<3\times10^{-9}$ & 2.200 \\
0.180 &  1.60518 & 1.31149 &  1.224 &   0.26127 & -0.65212 & 1.62518 & $<3\times10^{-9}$ & 1.840 \\
0.200 &  1.33901 & 1.45908 &  0.918 &   0.18653 & -0.80087 & 1.38273 & $<3\times10^{-9}$ & 1.600 \\
\hline
\end{tabular}%
}
\end{table}

\subsection{Geometric wall crossover}
\label{sec:wall-crossover}

The reduced-model continuation reaches the geometric criterion in
\cref{eq:wall-crossover-criterion} at
\begin{equation}
 \dlt_{\rm wall}=0.2492098.
 \label{eq:wall-crossover-values}
\end{equation}
At this point the radial shape state remains below the false-vacuum continuum,
\begin{equation}
 \Lam_{{\rm sh},0}\simeq0.8271,
 \qquad
 \Lam_{\rm f}\simeq1.0095,
 \qquad
 \Lam_{\rm f}-\Lam_{{\rm sh},0}\simeq0.1824.
 \label{eq:wall-crossover-gap}
\end{equation}
The geometric crossover is therefore not an eigenvalue crossing.  Both
angular threshold endpoints occur at smaller supercooling, whereas the
$\ell=0$ state remains localized until the half-height radius itself ceases
to define a distinct wall.  We do not continue the wall-centered diagnostics
beyond this point.

The finite-difference spectrum shows the expected second-order grid
convergence, and the Richardson-extrapolated $\ell=1$ translational zero mode
remains below $3\times10^{-9}$ in magnitude throughout the main scan.
Outer-boundary, grid-spacing, threshold-integration, and matching-radius
checks are collected in \cref{app:merger-convergence}.

\section{Discussion}
\label{sec:discussion}

The ordering of the angular mergers follows directly from the structure of
the radial operator.  The false-vacuum continuum threshold is independent of
$\ell$, whereas the centrifugal term raises the localized angular states, so
the $\ell=2$ member reaches the continuum before $\ell=1$.  The leading
thin-wall estimate is unusually accurate for $\ell=1$; for $\ell=2$,
higher-order finite-supercooling corrections are comparatively important,
consistent with the near cancellation in its quadratic coefficient
$9(10-\pi^2)$.  Both angular endpoints occur well before the geometric wall
crossover.  Spectral dissolution of an angular branch and loss of a
recognizable interface are therefore distinct phenomena.

The surviving $\ell=0$ state exhibits a different evolution.  Its continuum
gap remains positive while the normalized wave function becomes strongly
asymmetric: the planar overlap decreases, the radial probability centroid
moves toward the false-vacuum exterior, and the exterior localization length
grows relative to the interior scale.  The unequal tail lengths follow from
the bulk-curvature hierarchy $\Lam_{\rm f}<\Lam_{\rm t}$, while spherical
curvature and wall-core relaxation control the detailed profile.  These
wave-function diagnostics reveal information that is not encoded by the
eigenvalue alone and would be hidden in an aggregate functional determinant
unless the corresponding state were isolated.

The spectrum analyzed here is that of the Euclidean Hessian around the
critical bubble.  The subthreshold bound states and the square-integrable
threshold states at the merger points provide normalizable deformation
profiles of the nucleation saddle.  Beyond the $\ell=1,2$ endpoints, a
scattering or real-time description requires continuum wave packets or
resonance data rather than finite-box levels.  Microscopic matching will also
introduce temperature-dependent couplings, moving vacua, kinetic corrections,
and the remainder in \cref{eq:near-critical-decomposition}, which can shift
the numerical merger locations.  The quoted endpoint values are therefore
precise properties of the reduced tilted quartic.  The common false-vacuum
threshold, the centrifugal ordering of the mergers, and the separation
between spectral and geometric endpoints are the structural features exposed
by the analysis.

Natural extensions include above-threshold resonance calculations using
phase shifts, Jost functions, outgoing boundary conditions, or complex
scaling, as well as multi-field generalizations with mixed-channel
thresholds.  A separate Lorentzian treatment is required to determine how
the surviving radial deformation evolves after nucleation.

\section{Conclusions}
\label{sec:conclusion}

We have determined the partial-wave-dependent threshold endpoints of the
positive shape band of an $O(3)$ critical bubble and tracked the surviving
radial state through substantial wave-function deformation at finite
supercooling.  Threshold shooting gives
$\dlt_{{\rm merge},2}=0.0900472$ and
$\dlt_{{\rm merge},1}=0.1410162$.  At each endpoint the threshold state
remains square integrable, while finite-box calculations confirm that the
post-merger levels are continuum states rather than normalizable eigenstate
continuations.

The $\ell=0$ shape state follows a different path.  It remains below the
false-vacuum continuum through the wall-resolved branch and up to the
geometric crossover at $\dlt_{\rm wall}=0.2492098$, while its planar overlap,
radial centroid, and interior and exterior decay scales reveal strong
asymmetric deformation.  The established thin-wall coefficients are
recovered in the near-coexistence limit, providing an analytic and numerical
consistency check on the continuation.

The planar shape mode therefore develops two distinct finite-supercooling
fates: angular branches terminate at partial-wave-dependent continuum
thresholds, whereas the radial branch remains localized while deforming
asymmetrically.  Because the angular threshold endpoints precede the loss of
a recognizable wall, spectral dissolution and geometric wall crossover are
independent diagnostics.  Their separation provides a clean mode-resolved
benchmark for the finite-supercooling evolution of critical-bubble
excitations.

\bibliographystyle{ptephy}
\bibliography{references}

\appendix

\section{Exact maps between quartic parametrizations}
\label{app:quartic-maps}

The reduced tilted quartic used in the main text is an exact parametrization
of the standard one-component quartic family.  For reference, this appendix
collects the field, coordinate, and eigenvalue maps needed to compare our
conventions with commonly used parametrizations.

\subsection{Cubic--quartic parametrization}
\label{app:cubic-quartic-map}

After the field reflection $\phi_C=-\phi$, it takes the cubic form
used in modern thin-to-thick-wall studies,
\begin{equation}
 V_C(\phi_C)
 =\frac{1}{2}m^2\phi_C^2+\eta_C\phi_C^3
 +\frac{\lambda_C}{8}\phi_C^4,
 \label{eq:cubic-quartic-parametrization}
\end{equation}
with
\begin{equation}
 m^2=\frac{\lambda\vev^2}{2}(1-3\dlt),
 \qquad
 \eta_C=\frac{\lambda\vev}{2}(1-\dlt),
 \qquad
 \lambda_C=2\lambda.
 \label{eq:cubic-quartic-map}
\end{equation}
The thin-to-thick-wall parameter of Ref.~\cite{Matteini2025} is therefore
\begin{equation}
 \epsilon_\alpha
 =1-\frac{\lambda_Cm^2}{4\eta_C^2}
 =\frac{\dlt(1+\dlt)}{(1-\dlt)^2}.
 \label{eq:epsilon-alpha-map}
\end{equation}
This mapping makes explicit that existing calculations of the action and full
functional determinant for a general quartic potential apply to the same
underlying family.

\subsection{M\"unster--Rotsch parametrization}
\label{app:MR-map}

An affine field shift brings the model to the linearly tilted symmetric
double well used by M\"unster and Rotsch \cite{MuensterRotsch2000}.  A useful
exact map is
\begin{equation}
 \Phi_{\rm MR}=\frac{\vev}{2}(1-2s-\dlt),
 \qquad
 g_{\rm MR}=6\lambda,
 \qquad
 v_{\rm MR}=\frac{\vev}{2}\sqrt{1+3\dlt^2}.
 \label{eq:MR-field-map}
\end{equation}
In terms of their dimensionless asymmetry, radial coordinate, and Hessian
eigenvalue,
\begin{equation}
 \widetilde\eta
 =\frac{3\dlt(1-\dlt^2)}{(1+3\dlt^2)^{3/2}},
 \qquad
 \widetilde r=\sqrt{1+3\dlt^2}\,\rho,
 \qquad
 \Lam=(1+3\dlt^2)\,\omega.
 \label{eq:MR-spectral-map}
\end{equation}
These relations are used in \cref{sec:near-critical-operator} to translate
the known analytic thin-wall spectrum into our conventions.

\subsection{High-temperature parametrization}
\label{app:high-temperature-map}

For comparison with a commonly used high-temperature quartic representation,
introduce the dimensionless field $\varphi$ and parameter $q$ through
\begin{equation}
 U_{\rm HT}(\varphi;q)=\varphi^4-2\varphi^3+q\varphi^2,
 \qquad
 \varphi_{\rm t}=\frac{3+\sqrt{9-8q}}{4}.
 \label{eq:high-temperature-quartic-family}
\end{equation}
Defining $s=\varphi/\varphi_{\rm t}$ and
$\dlt=1-\varphi_{\rm t}^{-1}$ gives the identity
\begin{equation}
 U_{\rm HT}(\varphi;q)
 =\varphi_{\rm t}^4
 \left[s^2(1-s)^2-\dlt(3s^2-2s^3)\right],
 \qquad
 q=\frac{1-3\dlt}{(1-\dlt)^2}.
 \label{eq:q-delta-exact-map}
\end{equation}
The corresponding dimensionless radial coordinates and Hessian eigenvalues
are related by
\begin{equation}
 \rho=\frac{\varphi_{\rm t}}{\sqrt{2}}x_{\rm HT},
 \qquad
 \Lam_{\rm HT}=\frac{\varphi_{\rm t}^2}{2}\Lam.
 \label{eq:HT-coordinate-eigenvalue-map}
\end{equation}
This map provides an exact translation between the reduced variables used in
the main text and the standard high-temperature quartic family.

\section{Numerical diagnostics and threshold-shooting convergence}
\label{app:merger-convergence}

\subsection{Near-coexistence fit diagnostics}

\Cref{tab:fit-diagnostics} summarizes the residuals of the asymptotic fits and
the variation of the leading coefficient when the upper fit limit is changed
from $0.03$ to $0.02$ or $0.05$ while retaining the same polynomial basis.

\begin{table}[ht]
\centering
\caption{Diagnostics for the near-coexistence least-squares fits.  The central
fit uses the six points with $\dlt\leq0.03$.  Where an analytic coefficient is
known from the thin-wall spectrum, it is shown for comparison.  The last
column gives the range of the fitted leading coefficient when the upper fit
limit is changed to $0.02$, $0.03$, or $0.05$.}
\label{tab:fit-diagnostics}
\resizebox{\textwidth}{!}{%
\begin{tabular}{lrrrr}
\hline
Quantity fitted & Numerical coefficient & Analytic coefficient &
Maximum residual & Fit-window range \\
\hline
$R_{1/2}/\Lzero-1/(3\dlt)$
& $-0.9674$ & -- & $8.8\times10^{-8}$ & $[-0.96740,-0.96734]$ \\
$\Shat-1/(81\dlt^2)$
& $-0.10749$ & -- & $3.5\times10^{-7}$ & $[-0.107489,-0.107488]$ \\
$L_{\rm slope}/\Lzero-1$
& $5.317$ & -- & $2.2\times10^{-8}$ & $[5.316,5.318]$ \\
$\Lam_{{\rm tr},0}$
& $-18.000$ & $-18$ & $1.9\times10^{-8}$ & $[-18.000,-17.998]$ \\
$\Lam_{{\rm tr},2}$
& $36.00$ & $36$ & $1.1\times10^{-7}$ & $[36.000,36.011]$ \\
$\Lam_{{\rm sh},0}-3$
& $-52.82$ & $9(4-\pi^2)$ & $1.4\times10^{-7}$ & $[-52.83,-52.81]$ \\
$\Lam_{{\rm sh},1}-3$
& $-34.826$ & $9(6-\pi^2)$ & $1.5\times10^{-8}$ & $[-34.8264,-34.8250]$ \\
$\Lam_{{\rm sh},1}-\Lam_{{\rm sh},0}$
& $18.00$ & $18$ & $1.3\times10^{-7}$ & $[17.989,18.000]$ \\
$\Lam_{{\rm sh},2}-\Lam_{{\rm sh},0}$
& $54.0$ & $54$ & $1.2\times10^{-6}$ & $[54.00,54.12]$ \\
\hline
\end{tabular}%
}
\end{table}

\subsection{Finite-difference convergence}

The $\ell=1$ translational mode, the $\ell=0$ negative mode, and the
$\ell=0$ shape mode show the expected second-order grid convergence before
Richardson extrapolation.  The extrapolated translational zero mode remains
below $3\times10^{-9}$ in magnitude throughout the main scan.

\begin{figure}[ht]
 \centering
 \includegraphics[width=0.66\textwidth]{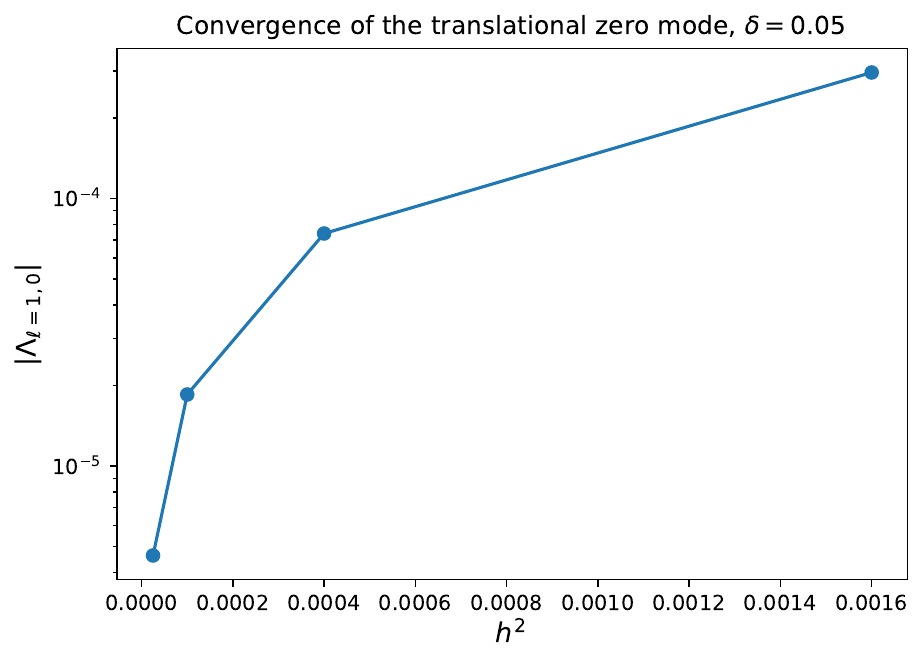}
 \caption{Grid convergence of the $\ell=1$ translational zero mode at
 $\dlt=0.05$.}
 \label{fig:zero-convergence}
\end{figure}

\begin{figure}[ht]
 \centering
 \includegraphics[width=0.66\textwidth]{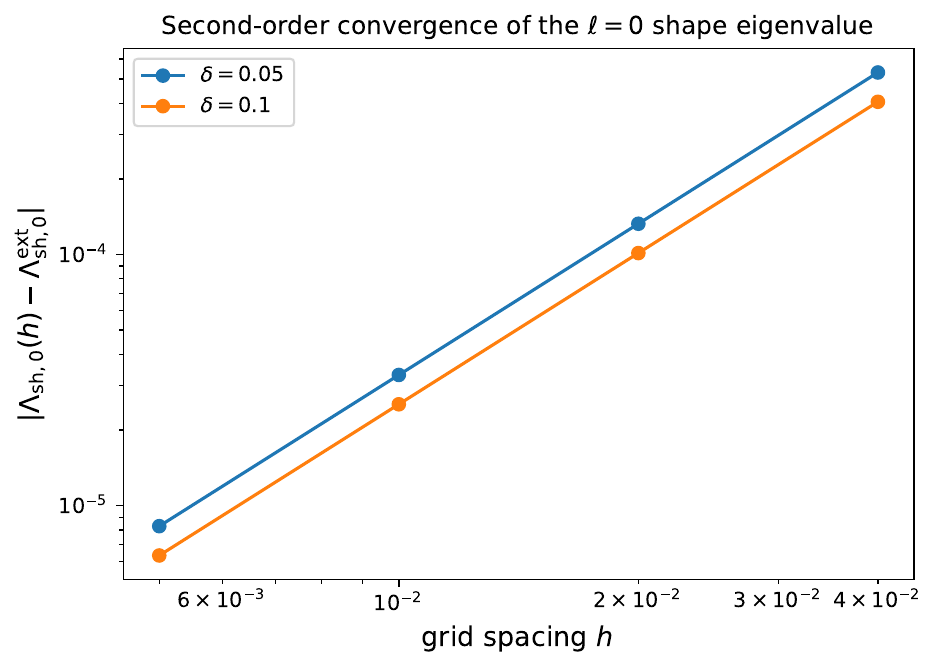}
 \caption{Absolute grid error of the $\ell=0$ shape eigenvalue relative to the
 Richardson-extrapolated value.  The nearly parallel slopes confirm
 second-order convergence at $\dlt=0.05$ and $0.10$.}
 \label{fig:shape-convergence}
\end{figure}

Outer-boundary tests varying both $\rho_{\rm BVP}$ and $\rho_{\rm box}$ change
the extrapolated $\ell=0$ shape eigenvalue by at most
$9.4\times10^{-10}$ at the tested points $\dlt=0.05$, $0.10$, and $0.20$.

\subsection{Threshold shooting and finite-box checks}

\Cref{tab:merger-step-convergence} shows the fixed-step convergence at
$\rho_{\rm m}=40$, and \cref{tab:merger-boundary-convergence} shows the
matching-radius dependence at $h=0.00125$.  The reference values use the
adaptive integration described in \cref{sec:threshold-shooting}.

\begin{table}[ht]
\centering
\caption{Step-size convergence of the threshold-shooting merger points.}
\label{tab:merger-step-convergence}
\begin{tabular}{ccccc}
\hline
$\ell$ & $h$ & $\dlt_{\rm merge}$ & shift from reference & threshold nodes \\
\hline
2 & 0.04000 & 0.0900471822 & $+1.84\times10^{-8}$ & 1 \\
2 & 0.01000 & 0.0900471639 & $+8.18\times10^{-11}$ & 1 \\
2 & 0.00250 & 0.09004716385 & $+3.59\times10^{-13}$ & 1 \\
2 & 0.00125 & 0.09004716385 & $+6.64\times10^{-14}$ & 1 \\
1 & 0.04000 & 0.1410162762 & $+4.55\times10^{-8}$ & 1 \\
1 & 0.01000 & 0.1410162309 & $+1.90\times10^{-10}$ & 1 \\
1 & 0.00250 & 0.14101623071 & $+7.35\times10^{-13}$ & 1 \\
1 & 0.00125 & 0.14101623071 & $-1.99\times10^{-14}$ & 1 \\
\hline
\end{tabular}
\end{table}

\begin{table}[ht]
\centering
\caption{Matching-radius convergence at $h=0.00125$.}
\label{tab:merger-boundary-convergence}
\begin{tabular}{cccc}
\hline
$\ell$ & $\rho_{\rm m}$ & $\dlt_{\rm merge}$ & shift from reference \\
\hline
2 & 12 & 0.0900471596 & $-4.22\times10^{-9}$ \\
2 & 15 & 0.09004716384 & $-8.46\times10^{-12}$ \\
2 & 20 & 0.09004716385 & $+6.60\times10^{-14}$ \\
2 & 60 & 0.09004716385 & $+6.64\times10^{-14}$ \\
1 & 12 & 0.1410162155 & $-1.52\times10^{-8}$ \\
1 & 15 & 0.14101623063 & $-8.38\times10^{-11}$ \\
1 & 20 & 0.14101623071 & $-3.79\times10^{-14}$ \\
1 & 60 & 0.14101623071 & $-2.00\times10^{-14}$ \\
\hline
\end{tabular}
\end{table}

The finite-box test provides an independent distinction between a bound state
and a continuum level.  Five thousandths below the merger, the extracted gaps
converge to $0.038372$ for $\ell=2$ and $0.009695$ for $\ell=1$.  Five
thousandths above the merger, the second Dirichlet level stays above
$\Lam_{\rm f}$ and approaches it as the box grows from $20$ to $60$.

\end{document}